\documentclass[aps,prd,twocolumn,floatfix,superscriptaddress,nofootinbib,10pt]{revtex4-2}

\usepackage[english]{babel}
\usepackage[T1]{fontenc}
\usepackage{amsmath}
\usepackage{amssymb}
\usepackage{graphicx}
\usepackage{xcolor}
\usepackage[colorlinks]{hyperref}
\usepackage{orcidlink}

\begin{document}

\title{Locally Rotationally Symmetric spacetimes of type II in \texorpdfstring{$f(\mathcal{Q})$}{} gravity}
\author{Fabrizio Esposito \orcidlink{0000-0001-6883-152X}}
\email{fabrizio.esposito01@edu.unige.it}
\affiliation{DIME, Università di Genova, Via all'Opera Pia 15, 16145 Genova, Italy}
\affiliation{INFN Sezione di Genova, Via Dodecaneso 33, 16146 Genova, Italy}
\author{Sante Carloni \orcidlink{0000-0003-2373-2653}}
\email{sante.carloni@unige.it}
\affiliation{DIME, Università di Genova, Via all'Opera Pia 15, 16145 Genova, Italy}
\affiliation{INFN Sezione di Genova, Via Dodecaneso 33, 16146 Genova, Italy}
\affiliation{Institute of Theoretical Physics, Faculty of Mathematics and Physics,
Charles University, Prague, V Hole{\v s}ovi{\v c}k{\' a}ch 2, 180 00 Prague 8, Czech Republic}
\author{Stefano Vignolo \orcidlink{0000-0003-2926-0650}}
\email{stefano.vignolo@unige.it}
\affiliation{DIME, Università di Genova, Via all'Opera Pia 15, 16145 Genova, Italy}
\affiliation{INFN Sezione di Genova, Via Dodecaneso 33, 16146 Genova, Italy}

\begin{abstract}
We investigate the $1+1+2$ covariant formalism in the presence of nonmetricity. Focusing on Locally Rotationally Symmetric spacetimes, we show how nonmetricity affects all the kinematic quantities involved in the covariant $1+1+2$ decomposition. We apply the resulting geometrical framework to study both homogeneous solutions and static spherically symmetric solutions in the context of $f(\mathcal{Q})$ gravity. We obtain sufficient conditions for homogeneous solutions with flat spatial hypersurfaces and Schwarzschild-de Sitter type solutions in the $1+1+2$ formalism. We also explore an elementary gravastar solution utilizing covariant junction conditions.
\end{abstract}

\date{\today}

\maketitle

\section{introduction}
General Relativity (GR) is a geometric theory of gravity that describes the gravitational field in terms of the metric tensor $g_{ab}$ and relates the matter distribution to the curvature of the spacetime, $R^{a}{}_{bcd}$. In addition to GR, and since its very first formulation, other metric theories of gravity were developed to extend the geometrization of fundamental forces and, later, to address the limitations of GR on quantum, astrophysical, and cosmological scales.  These theories can contain additional fields linked to the gravitational interaction or even have field equations containing higher-than-second derivatives (of the metric tensor). Well-known examples are $f(R)$ gravity, a fourth-order theory \cite{Clifton_2012,CAPOZZIELLO2011167,Sotiriou_2010,Nojiri:2010wj,Nojiri:2017ncd}, or Hordenski theories (see e.g. \cite{Kobayashi:2019hrl}). 

Further alternative theories are the ones based on torsion $T_{ab}{}^{c}$ and nonmetricity $Q_{abc}$ tensors, which, together with the curvature, express the properties of the affine connection defined on the spacetime. Among them, undoubtedly, the most studied is the so-called Teleparallel Gravity (TG). TG represents a class of gravitational gauge theory that assumes spacetime is flat and describes gravitation through the torsion \cite{Kr_k_2019}. Recently, much attention has been devoted to another class of teleparallel theories of gravity: Symmetric Teleparallel Gravity (STG). STG is a class of theories in which one assumes a torsion-free and curvature-free connection, but with nonmetricity that differs from zero \cite{Nester:1998mp,Adak:2005cd,Adak:2008gd,Conroy:2017yln}. Although teleparallel theories are geometric descriptions of gravity different from GR, in some remarkable cases, they have field equations equivalent to those of GR. These are called Teleparallel Equivalent of General Relativity (TEGR) and Symmetric Teleparallel Equivalent of General Relativity (STEGR). TG and STG have also been modified, generating what are now commonly known as $f(T)$-gravity and $f(\mathcal{Q})$-gravity \cite{BeltranJimenez:2017tkd,Bahamonde_2023,Cai:2015emx,Jim_nez_2020,Vignolo:2021frk,Esposito:2021ect,Esposito:2022omp,DAmbrosio:2021pnd,Gadbail:2022jco,Albuquerque:2022eac,Capozziello:2022wgl,Khyllep:2022spx,Agrawal:2022vdg,Narawade:2022jeg,Anagnostopoulos:2022gej,Xu:2019sbp,Yang:2021fjy,Iosifidis:2018diy,Capozziello:2022zzh,Paliathanasis:2023nkb,Heisenberg:2023lru}. 

This last class of theories, in particular, has been recently studied in great detail in cosmological and astrophysical settings. In $f(\mathcal{Q})$ gravity, a way to simplify the calculations is to choose the so-called ``coincident gauge''. In this specific gauge, the full connection is assumed to be identically zero so that the nonmetricity tensor is simply expressed by the partial derivative of the metric tensor \cite{BeltranJimenez:2017tkd}. However, the coincident gauge has some limitations. For example, in the context of astrophysical relevant spacetimes, such as the spherically symmetric ones we are interested in, the coincident gauge is incompatible with the usual spherical coordinates used to obtain the standard diagonal form of the metric. We refer to \cite{Bahamonde:2022zgj} for a detailed analysis of spherically symmetric spacetimes by using the coincident gauge. Several other approaches, alternative to the coincident gauge, have been proposed both in cosmology and astrophysics. For example, the total connection is made to coincide with the Levi-Civita one in the gravity-free case \cite{Zhao:2021zab, Lin:2021uqa, Calza:2022mwt}. Also, restrictions on the connection itself and exact solutions are found via the imposition of connection invariance under $SO(3)$ group transformations, or the invariance under both $SO(3)$ group and translation transformations \cite{Hohmann:2021ast,Dimakis:2023uib,DAmbrosio:2021pnd,Dimakis:2022rkd,DAmbrosio:2021zpm,Paliathanasis:2023ngs,Bahamonde:2022esv,Bahamonde:2022kwg,Dimakis:2024fan,Chakraborty:2025qlv}.

This paper aims to provide a new framework to study cosmological and astrophysical relevant metrics in $f(\mathcal{Q})$ gravity with a formalism that precludes the assumptions on the gauge and, therefore, treats the nonmetricity as a variable. More specifically, our analysis will be performed on both spatially homogeneous and static Locally Rotationally Symmetric (LRS) of type II spacetimes \cite{Stewart:1967tz} using the $1+1+2$ covariant formalism, a natural extension of $1+3$ formalism \cite{vanElst:1995eg,Clarkson:2002jz,Clarkson:2007yp,Luz:2019frs,Naidu:2022igk} that we used in \cite{Esposito:2022omp}. In this approach, in addition to splitting spacetime in a preferred temporal direction and spatial hypersurfaces, the space is further split into a preferred spatial direction and surfaces orthogonal to it. This makes the formalism well adapted to describe LRS spacetimes, which have a local preferred spatial direction corresponding to the local axis of symmetry.
We will see, in particular, that combining the formalism with the symmetries of LRS spacetimes allows us to build a complete set of scalar quantities that can describe the structure of homogeneous or static spherically symmetric spacetimes. Additionally, we will derive the physical consequences of nonmetricity within this framework.

The paper is organized as follows. In Sec. \ref{sec:geometric_framework}, we introduce the geometric framework performing a detailed analysis of the $1+1+2$ formalism when both curvature and nonmetricity tensors are considered. Section \ref{sec:LRS_spacetime} is dedicated to the description of constraints induced by the assumptions of the local rotational symmetry. We also show how these constraints allow the decomposition of the nonmetricity tensor in a set of scalar quantities. Using these results, in Sec. \ref{sec:f_Q_gravity}, we derive the 1+1+2 equations for $f(\mathcal{Q})$ gravity in LRS spacetimes. In Sec. \ref{sec:hom_spacetime}, we restrict to the homogeneous case and consider Friedmann–Lemaître–Robertson–Walker (FLRW) type metrics, whereas Sec. \ref{sec:static_spacetime} focuses on the static and spherical symmetric case and Schwarzschild-de Sitter type spacetimes. We also analyze the junction conditions when a Gravastar-like model is considered. Finally, we discuss the results in Sec. \ref{sec:conclusions}.

Throughout the paper natural units ($c=8\pi G=1$) and the metric signature $(-,+,+,+)$ are used.

\section{Geometric framework}\label{sec:geometric_framework}

We consider a $4$-dimensional manifold endowed with a metric tensor $g_{ab}$ and an affine connection $\Gamma_{ab}{}^{c}$, which is assumed torsionless but non-metric compatible. We denote by $\nabla$ the covariant derivative induced by $\Gamma_{ab}{}^{c}$,
and by
\begin{equation}\label{eq:def_nonmetricity}
    Q_{cab}=\nabla_{c}g_{ab}.
\end{equation}
the corresponding nonmetricity tensor.

According to the above assumptions, the connection $\Gamma_{ab}{}^{c}$ can be decomposed as
\begin{equation}\label{eq:def_connection}
\Gamma_{ab}{}^{c} = \tilde{\Gamma}_{ab}{}^{c} + L_{ab}{}^{c} ,
\end{equation}
where
\begin{equation}\label{eq:def_levicivita}
\tilde{\Gamma}_{ab}{}^{c} = \frac{1}{2} g^{cd} \left( \partial_{a}g_{bd} + \partial_{b}g_{ad} - \partial_{d}g_{ab} \right),
\end{equation}
is the Levi-Civita connection induced by the metric tensor $g_{ab}$, and 
\begin{equation}\label{eq:def_disformation}
L_{ab}{}^{c} = \frac{1}{2} \left( Q^{c}{}_{ab} - Q_{a}{}^{c}{}_{b} - Q_{b}{}^{c}{}_{a} \right)
\end{equation}
is the disformation tensor. Henceforth, we will denote by a tilde all quantities related to the Levi-Civita connection \eqref{eq:def_levicivita}.

The Riemann tensor $R^{a}{}_{bcd}$ induced by the connection $\Gamma_{ab}{}^{c}$ can be expressed as
\begin{equation}\label{eq:def_riemann_1}
\begin{split}
    R^{a}{}_{bcd} =& \partial_{c}\Gamma_{db}{}^{a} - \partial_{d}\Gamma_{cb}{}^{a} + \Gamma_{cp}{}^{a}\Gamma_{db}{}^{p} - \Gamma_{dp}{}^{a}\Gamma_{cb}{}^{p} =\\
    =& \tilde{R}^{a}{}_{bcd} + \tilde{\nabla}_{c}L_{db}{}^{a} - \tilde{\nabla}_{d}L_{cb}{}^{a} + L_{cp}{}^{a}L_{db}{}^{p} +\\
    &- L_{dp}{}^{a}L_{cb}{}^{p},
\end{split}
\end{equation}
with
\begin{equation}
     \tilde{R}^{a}{}_{bcd} = \partial_{c}\tilde{\Gamma}_{db}{}^{a} - \partial_{d}\tilde{\Gamma}_{cb}{}^{a} + \tilde{\Gamma}_{cp}{}^{a}\tilde{\Gamma}_{db}{}^{p} - \tilde{\Gamma}_{dp}{}^{a}\tilde{\Gamma}_{cb}{}^{p},
\end{equation}
and it satisfies the following identities:
\begin{itemize}\label{eq:riemann_properties}
    \item Antisymmetry in the last two indices,
    \begin{equation}
       R^{a}{}_{bcd} = -R^{a}{}_{bdc};
    \end{equation}
    \item First Bianchi identity,
    \begin{equation}\label{eq:first_bianchi_identity}
       R^{a}{}_{[bcd]}=0;
    \end{equation}
    \item Second Bianchi identity,
    \begin{equation}\label{eq:second_bianchi_identity}
        \nabla_{[e|}R^{a}{}_{b|cd]}=0.
    \end{equation}
\end{itemize}
Moreover, the Riemann tensor \eqref{eq:def_riemann_1} has three independent traces:
\begin{equation}\label{eq:riemann_traces}
\begin{gathered}
    R_{bd} = R^{a}{}_{bad}, \qquad \bar{R}_{ad} = R_{a}{}^{c}{}_{cd}, \\ \check{R}_{cd} = R^{a}{}_{acd} = - \partial_{[c}Q_{d]a}{}^{a}.
\end{gathered}
\end{equation}
The contraction of Eq. \eqref{eq:riemann_traces} with the metric gives us the Ricci scalar, 
\begin{equation}\label{eq:def_ricci_scalar}
    R= g^{ab}R_{ab} = - g^{ab}\bar{R}_{ab}.
\end{equation}
From Eq. \eqref{eq:def_riemann_1}, it is easily seen that the Ricci scalar can be expressed as
\begin{equation}
    R = \tilde{R} + \tilde{\nabla}_{a}L_{b}{}^{ba}- \tilde{\nabla}_{b}L_{a}{}^{ba} + \mathcal{Q},
\end{equation}
where the quantity
\begin{equation}\label{eq:def_nonmetricity_scalar}
\begin{split}
    \mathcal{Q} &= L_{ap}{}^{a}L_{b}{}^{bp} - L_{bp}{}^{a}L_{a}{}^{bp} =\\
    =& \frac{1}{4}Q_{cab}Q^{cab} - \frac{1}{2}Q_{cab}Q^{abc} - \frac{1}{4} Q_{ab}{}^{b}Q^{a}{}_{b}{}^{b} + \frac{1}{2}Q_{ab}{}^{b}Q_{ba}{}^{b},
\end{split}    
\end{equation}
is the so-called nonmetricity scalar. The Riemann tensor can also be written as
\begin{equation}\label{eq:def_irr_dec_riemann}
    R_{abcd} = C_{abcd} + \sum_{n=2}^{6} W^{(n)}_{abcd} + \sum_{n=2}^{5} Z^{(n)}_{abcd},
\end{equation}
where
\begin{equation}\label{eq:def_weyl_tensor}
    C_{abcd} = W^{(1)}_{abcd} + Z^{(1)}_{abcd}
\end{equation}
is the Weyl tensor. Further details are provided in Appendix \ref{appendix_weyl}.


\subsection{\texorpdfstring{$1+1+2$ decomposition}{}}

Let us consider a congruence of timelike curves with tangent vector field $u$:
\begin{equation}
    g_{ab}u^{a}u^{b}=-1.
\end{equation} 
At any point, we can identify a $3$-dimensional subspace of the tangent bundle orthogonal to $u^{a}$. The projector on this spatial subspace is defined as
\begin{equation}\label{eq:def_3_metric}
    h_{ab} = g_{ab} + u_{a}u_{b},
\end{equation}
and satisfies the properties
\begin{equation}
    h^{a}{}_{c}h^{c}{}_{b}=h^{a}{}_{b}, \qquad h_{ab}u^{b}=0, \qquad  h_{a}{}^{a} = 3.
\end{equation}
The $1+1+2$ decomposition is implemented by introducing a congruence of spacelike curves with a tangent vector field $e$ everywhere orthogonal to $u$:
\begin{equation}
    g_{ab}e^{a}e^{b} = 1, \qquad g_{ab}u^{a}e^{b}=0.
\end{equation}
This allows us to split the spatial subspace into a preferred direction parallel to $e$ and a $2$-dimensional subspace, called ``sheet'', orthogonal to $e$. The projection tensor onto the sheet is given by
\begin{equation}\label{eq:def_2_metric}
    N_{ab} = h_{ab} - e_{a}e_{b},
\end{equation}
with
\begin{equation}
    N^{a}{}_{c}N^{c}{}_{b} = N^{a}{}_{b}, \quad N_{ab}u^{b} = N_{ab}e^{b} =0, \quad N_{a}{}^{a} = 2.
\end{equation}
The norm conservation of the vectors $u$ and $e$ along both the congruences and with respect to the covariant derivative induced by the full connection $\Gamma_{ab}{}^{c}$, as well as the preservation of their orthogonality, impose restrictions on the nonmetricity tensor $Q_{cab}$:
\begin{equation}\label{eq:nonmetricity_cond_1}
\begin{gathered}
    u^{c}\nabla_{c}\left(g_{ab}u^{a}u^{b}\right) = Q_{cab}u^{c}u^{a}u^{b} + 2u_{b}u^{c}\nabla_{c}u^{b} = 0 \\ \longrightarrow \quad Q_{cab}u^{c}u^{a}u^{b} = 2u^{b}u^{c}\nabla_{c}u_{b},
\end{gathered}
\end{equation}
\begin{equation}\label{eq:nonmetricity_cond_2}
\begin{gathered}
    e^{c}\nabla_{c}\left(g_{ab}e^{a}e^{b}\right) = Q_{cab}e^{c}e^{a}e^{b} + 2e_{b}e^{c}\nabla_{c}e^{b} = 0 \\ \longrightarrow \quad Q_{cab}e^{c}e^{a}e^{b} = 2e^{b}e^{c}\nabla_{c}e_{b},
\end{gathered}
\end{equation}
\begin{equation}\label{eq:nonmetricity_cond_3}
\begin{gathered}
    e^{c}\nabla_{c}\left(g_{ab}u^{a}u^{b}\right) = Q_{cab}e^{c}u^{a}u^{b} + 2u_{b}e^{c}\nabla_{c}u^{b} = 0 \\ 
    \longrightarrow \quad Q_{cab}e^{c}u^{a}u^{b} = 2u^{b}e^{c}\nabla_{c}u_{b},
\end{gathered}
\end{equation}
\begin{equation}\label{eq:nonmetricity_cond_4}
\begin{gathered}
    u^{c}\nabla_{c}\left(g_{ab}e^{a}e^{b}\right) = Q_{cab}u^{c}e^{a}e^{b} + 2e_{b}u^{c}\nabla_{c}e^{b} = 0 \\
    \longrightarrow \quad Q_{cab}u^{c}e^{a}e^{b} = 2e^{b}u^{c}\nabla_{c}e_{b},
\end{gathered}
\end{equation}
\begin{equation}\label{eq:nonmetricity_cond_5}
\begin{gathered}
    u^{c}\nabla_{c}\left(g_{ab}u^{a}e^{b}\right) = Q_{cab}u^{c}u^{a}e^{b} + g_{ab}u^{c}\nabla_{c}\left(u^{a}e^{b}\right) = 0 \\
    \longrightarrow \quad Q_{cab}u^{c}u^{a}e^{b} =  u^{b}u^{c}\nabla_{c}e_{b} + e^{b}u^{c}\nabla_{c}u_{b},
\end{gathered}
\end{equation}
\begin{equation}\label{eq:nonmetricity_cond_6}
\begin{gathered}
    e^{c}\nabla_{c}\left(g_{ab}u^{a}e^{b}\right) = Q_{cab}e^{c}u^{a}e^{b} + g_{ab}e^{c}\nabla_{c}\left(u^{a}e^{b}\right) = 0 \\
    \longrightarrow \quad Q_{cab}e^{c}u^{a}e^{b} =  u^{b}e^{c}\nabla_{c}e_{b} + e^{b}e^{c}\nabla_{c}u_{b}.
\end{gathered}
\end{equation}
Making use of $h_{ab}$ and $N_{ab}$, we can define the projected symmetric trace-free (PSTF) part of a tensor for $u^{a}$ and $e^{a}$, respectively:
\begin{equation}\label{eq:def_PSTF_h}
    \psi_{\langle ab \rangle} =  \left[h_{(a}{}^{c}h_{b)}{}^{d} - \frac{1}{3}h_{ab}h^{cd}\right]\psi_{cd},
\end{equation}
and
\begin{equation}\label{eq:def_PSTF_N}
    \psi_{\{ ab \}} =  \left[N_{(a}{}^{c}N_{b)}{}^{d} - \frac{1}{2}N_{ab}N^{cd}\right]\psi_{cd},
\end{equation}
where $\psi_{ab}$ is a generic second rank covariant tensor.
It is also useful to introduce the volume elements derived from the Levi-Civita tensor $\epsilon_{abcd}$,
\begin{equation}\label{eq:def_volume_element}
    \epsilon_{abc} = \epsilon_{abcd}u^{d}, \qquad \qquad \epsilon_{ab} = \epsilon_{abc}e^{c},
\end{equation}
which are characterized by the following properties:
\begin{equation}
    \epsilon_{abc}=\epsilon_{[abc]}, \qquad \qquad \epsilon_{ab}=\epsilon_{[ab]},
\end{equation}
\begin{equation}
    \epsilon_{abc}u^{c} = 0, \quad \epsilon_{abc}\epsilon^{dec} = 2 h_{[a}{}^{d}h_{b]}{}^{e}, \quad \epsilon_{abc}\epsilon^{abd} =  2 h_{c}{}^{d}
\end{equation}
\begin{equation}
\begin{gathered}
    \epsilon_{ab}e^{b} = \epsilon_{ab}u^{b} = 0, \qquad \qquad \epsilon_{a}{}^{c}\epsilon_{cb} = N_{ab}, \\
    \epsilon_{abc} = e_{a}\epsilon_{bc} - e_{b}\epsilon_{ac} + e_{c}\epsilon_{ab}.
\end{gathered}
\end{equation}

\subsubsection{kinematic quantities}
Given a generic tensor $\psi^{a\cdot\cdot\cdot}{}_{b\cdot\cdot\cdot}$, there are several kinds of covariant derivative we can define in the $1+1+2$ formalism. In particular:
\begin{itemize}
    \item the time derivative, the covariant derivative along $u^{a}$,
    \begin{equation}\label{eq:def_time_derivative}
        \dot{\psi}^{a\cdot\cdot\cdot}{}_{b\cdot\cdot\cdot} = u^{c}\nabla_{c}\psi^{a\cdot\cdot\cdot}{}_{b\cdot\cdot\cdot};
    \end{equation}
    \item the spatial derivative, the covariant derivative projected onto the $3$-dimensional subspace orthogonal to $u^{a}$,
    \begin{equation}\label{eq:def_spatial_derivative}
        D_{c}\psi^{a\cdot\cdot\cdot}{}_{b\cdot\cdot\cdot} = h_{c}{}^{d} h^{a}{}_{e}\cdot\cdot\cdot h_{b}{}^{f}\cdot\cdot\cdot \nabla_{d}\psi^{e\cdot\cdot\cdot}{}_{f\cdot\cdot\cdot};
    \end{equation}
    \item the hat derivative, the covariant spatial derivative along $e^{a}$,
    \begin{equation}\label{eq:def_hat_derivative}
        \hat{\psi}^{a\cdot\cdot\cdot}{}_{b\cdot\cdot\cdot} = e^{c}D_{c}\psi^{a\cdot\cdot\cdot}{}_{b\cdot\cdot\cdot};
    \end{equation}
    \item the $\delta$-derivative, the covariant spatial derivative projected onto the sheet orthogonal to $e^{a}$,
    \begin{equation}\label{eq:def_delta_derivative}
        \delta_{c}\psi^{a\cdot\cdot\cdot}{}_{b\cdot\cdot\cdot} = N_{c}{}^{d} N^{a}{}_{e}\cdot\cdot\cdot N_{b}{}^{f}\cdot\cdot\cdot D_{d}\psi^{e\cdot\cdot\cdot}{}_{f\cdot\cdot\cdot}.
    \end{equation}
\end{itemize}
We can now decompose the covariant derivative of the $4$-velocity,
\begin{equation}\label{eq:def_covarinat_derivative_u}
\begin{split}
    \nabla_{a}u_{b} =& D_{a}u_{b} - u_{a}\left(\mathcal{A}_{b} + e_{b}\mathcal{A}^{(u)}\right) - \frac{1}{2}u_{b}N_{a}{}^{c}Q_{cde}u^{d}u^{e} +\\ 
    &- \frac{1}{2}u_{a}u_{b}Q_{cde}u^{c}u^{d}u^{e} - \frac{1}{2}u_{b}e_{a}Q_{cde}e^{c}u^{d}u^{e},
\end{split}
\end{equation}
with
\begin{equation}\label{eq:def_acceleration_u}
     \mathcal{A}^{(u)} = e^{a}u^{b}\nabla_{b}u_{a}, \qquad \text{and} \qquad \mathcal{A}_{b} = N_{b}{}^{a}u^{c}\nabla_{c}u_{a},
\end{equation}
the projections of the temporal acceleration of $u^{a}$ along $e^{a}$ and onto the sheet, respectively.

Moreover, the spatial derivative of $u_a$ can be expressed as
\begin{equation}\label{eq:def_spatial_derivative_u}
    D_{a}u_{b} = \frac{1}{3}h_{ab}\theta + \sigma_{ab} + \omega_{ab},
\end{equation}
where
\begin{equation}\label{eq:def_theta}
    \theta = h^{ab}D_{a}u_{b} 
\end{equation}
represents the expansion of the time congruence,
\begin{equation}\label{eq:def_shear_u}
\begin{split}
    \sigma_{ab} =& \left[h_{(a}{}^{c}h_{b)}{}^{d} - \frac{1}{3}h_{ab}h^{cd}\right]D_{c}u_{d} =\\
    =& \Sigma_{ab} + 2 \Sigma_{(a}e_{b)} + \Sigma \left(e_{a}e_{b} - \frac{1}{2}N_{ab} \right)
\end{split}
\end{equation}
represents the distortion, and it is called ``shear'' tensor, while
\begin{equation}
\begin{gathered}
    \Sigma_{ab} = \sigma_{\{ab\}}, \qquad \qquad \Sigma_{a} = N_{a}{}^{c}e^{b}\sigma_{cb}, \\
    \Sigma = e^{a}e^{b}\sigma_{ab} = - N^{ab}\sigma_{ab}
\end{gathered}
\end{equation}
are the tensor, vector, and scalar parts of the shear. Finally, the quantity
\begin{equation}\label{eq:def_vorticity_u}
    \omega_{ab} = D_{[a}u_{b]} = \epsilon_{ab}\Omega + \epsilon_{abc}\Omega^{c}
\end{equation}
is the ``vorticity'' tensor of the time congruence, with
\begin{equation}
    \Omega_{c} = \frac{1}{2}N_{cd}\epsilon^{abd}D_{a}u_{b}, \qquad \Omega = \frac{1}{2}\epsilon^{ab}N_{a}{}^{c}N_{b}{}^{d}D_{c}u_{d} .
\end{equation}
Similarly, the covariant derivative of $e_{a}$ can be expressed as
\begin{equation}\label{eq:def_covariant derivative_e}
\begin{split}
    \nabla_{a}e_{b} =& D_{a}e_{b} - u_{a}\alpha_{b} - \mathcal{A}^{(e)}u_{a}u_{b} +\\ 
    &+ \left( \frac{1}{3}\theta + \Sigma \right)e_{a}u_{b} - \mathcal{B}^{(e)}_{a}u_{b} +\\
    &- \frac{1}{2} u_{a}e_{b}Q_{cdf}u^{c}e^{d}e^{f} - e_{a}u_{b}Q_{cdf}e^{c}u^{d}e^{f},
\end{split}
\end{equation}
with
\begin{equation}
    \mathcal{B}^{(e)}_{a} = u^{d}N_{a}{}^{c}\nabla_{c}e_{d}.
\end{equation}
The spatial derivative of $e_a$ is given by
\begin{equation}\label{eq:def_spatial_derivative_e}
\begin{split}
    D_{a}e_{b} =& \frac{1}{2}N_{ab}\phi + \zeta_{ab} + \epsilon_{ab}\xi + e_{a}\mathrm{a}_{b} +\\
    &+ \frac{1}{2}N_{a}{}^{c}e_{b}Q_{cdf}e^{d}e^{f} + \frac{1}{2}e_{a}e_{b} Q_{cdf}e^{c}e^{d}e^{f},
\end{split}
\end{equation}
where
\begin{equation}\label{eq:def_phi_shear_vorticity_e}
    \phi = N^{ab}\delta_{a}e_{b}, \qquad \zeta_{ab} = \delta_{\{a}e_{b\}}, \qquad \xi = \frac{1}{2}\epsilon^{ab}\delta_{a}e_{b} 
\end{equation}
are the analogous expansion, shear, and vorticity for the spatial vector $e_{a}$. The scalar 
\begin{equation}\label{eq:def_acceleration_e}
    \mathcal{A}^{(e)} = -u^{b}u^{a}\nabla_{a}e_{b}
\end{equation}
is the projection of the temporal acceleration of $e^{a}$ along $u^{a}$, the vector 
\begin{equation}
    \alpha_{b} = u^{c}N_{b}{}^{d}\nabla_{c}e_{d}
\end{equation}
is the projection of the temporal acceleration of $e^{a}$ onto the sheet, and
\begin{equation}
    \mathrm{a}_{b} = e^{c}N_{b}{}^{d}\nabla_{c}e_{d}
\end{equation}
is the projection of the spatial acceleration onto the sheet.

It is worth noticing, and also useful for the definition of LRS spacetimes, that the tensors $\omega_{ab}$ and $\xi$ are equal to their Levi-Civita counterparts since we have torsion-free connections:
\begin{equation}
    \omega_{ab} = \tilde{\omega}_{ab}, \qquad \xi = \tilde{\xi}.
\end{equation}

\subsubsection{Magnetic and Electric parts of the Weyl tensor}

Making use of the Weyl tensor \eqref{eq:def_weyl_tensor}, we introduce six new tensors:
\begin{equation}\label{eq:def_magnetic_weyl}
\begin{gathered}
    H_{cd} = \frac{1}{2}\epsilon_{c}{}^{ab}u^{e}C_{abde}, \\
    \bar{H}_{ab} = \frac{1}{2}\epsilon_{a}{}^{cd} u^{e} C_{becd} = H_{ab} + \frac{1}{2}\epsilon_{a}{}^{cd}u^{e}Z^{(1)}{}_{becd}, \\
    \check{H}_{ab} = \frac{1}{2}\epsilon_{a}{}^{cd}u^{e}C_{ebcd} = -H_{ab} + \frac{1}{2}\epsilon_{a}{}^{cd}u^{e}Z^{(1)}{}_{ebcd},
\end{gathered}
\end{equation}
\begin{equation}\label{eq:def_electric_weyl}
\begin{gathered}
    E_{ac} = C_{abcd}u^{b}u^{d}, \qquad \qquad \bar{E}_{bc} = C_{abcd}u^{a}u^{d}, \\
    \check{E}_{cd} = C_{abcd}u^{a}u^{b},
\end{gathered}
\end{equation}
which are the ``magnetic'' and ``electric'' parts of the Weyl tensor, respectively. These tensors satisfy the following properties:
\begin{equation}
\begin{gathered}
    H_{cd} = h_{c}{}^{a}h_{d}{}^{b}H_{ab}, \quad H_{cd}=H_{dc}, \\
    \quad H^{a}{}_{a} = \bar{H}^{a}{}_{a} = \check{H}^{a}{}_{a} = 0,
\end{gathered} 
\end{equation}
\begin{equation}
    \check{E}_{cd} = - \check{E}_{dc}, \qquad E^{a}{}_{a} = \bar{E}^{a}{}_{a} = \check{E}^{a}{}_{a} = 0 .
\end{equation}

The magnetic and electric parts can be decomposed as well:
\begin{equation}\label{eq:dec_magnetic_weyl_1}
    H_{ab} = \mathcal{H}_{ab} + \mathcal{H}_{a}e_{b} + e_{a}\mathcal{H}_{b} + \left(e_{a}e_{b} - \frac{1}{2}N_{ab}\right) \mathcal{H} ,
\end{equation}
\begin{equation}\label{eq:dec_magnetic_weyl_2}
\begin{split}
    \bar{H}_{ab} =& \bar{\mathcal{H}}_{ab}  + \bar{\mathcal{H}}_{a}e_{b}  + e_{a} \mathbf{H}_{b} - \check{\bar{\mathcal{H}}}_{a}u_{b} + \epsilon_{ab}\bar{\mathbb{H}} +\\
    &- e_{a}u_{b}\check{\bar{\mathcal{H}}} + \left(e_{a}e_{b} - \frac{1}{2}N_{ab}\right) \bar{\mathcal{H}},
\end{split}
\end{equation}
\begin{equation}\label{eq:dec_magnetic_weyl_3}
\begin{split}
    \check{H}_{ab} =& \check{\mathcal{H}}_{ab}  + \check{\mathcal{H}}_{a}e_{b}  + e_{a} \check{\mathbf{H}}_{b} - \mathfrak{H}_{a}u_{b} + \epsilon_{ab}\check{\mathbb{H}} +\\
    &- e_{a}u_{b}\mathfrak{H} + \left(e_{a}e_{b} - \frac{1}{2}N_{ab}\right) \check{\mathcal{H}},
\end{split}
\end{equation}
\begin{equation}\label{eq:dec_electric_weyl_1}
\begin{split}
    E_{ab} =& \mathcal{E}_{ab} + \mathcal{E}_{a}e_{b} + e_{a} \mathbf{E}_{b} - u_{a} \check{\mathcal{E}}_{b} + \epsilon_{ab}\mathbb{E} +\\
    &- u_{a}e_{b} \check{\mathcal{E}} + \left(e_{a}e_{b} - \frac{1}{2}N_{ab}\right) \mathcal{E},
\end{split}
\end{equation}
\begin{equation}\label{eq:dec_electric_weyl_2}
\begin{split}
    \bar{E}_{ab} =& \bar{\mathcal{E}}_{ab} + \bar{\mathcal{E}}_{a}e_{b} + e_{a}\bar{\mathbf{E}}_{b} - u_{a} \check{\bar{\mathcal{E}}}_{b} + \epsilon_{ab} \bar{\mathbb{E}} +\\ &-u_{a}e_{b}\check{\bar{\mathcal{E}}} + \left(e_{a}e_{b} - \frac{1}{2}N_{ab}\right) \bar{\mathcal{E}},
\end{split}
\end{equation}
\begin{equation}\label{eq:dec_electric_weyl_3}
    \check{E}_{ab} = \mathfrak{E}_{ab} + 2\mathfrak{E}_{[a}e_{b]} - 2\check{\mathfrak{E}}_{[a}u_{b]} - 2 \mathfrak{E}u_{[a}e_{b]} + \epsilon_{ab}\check{\mathfrak{E}}.
\end{equation}
The Eqs. \eqref{eq:dec_magnetic_weyl_1}-\eqref{eq:dec_electric_weyl_3} involve tensor, vector, and scalar quantities which are defined in Appendix \ref{appendix_magnetic_eletric_weyl}.

\subsubsection{Gauss Relation}
Given a generic spatial vector field $v^{b}$, we can define the $3$-dimensional spatial Riemann tensor, $\, ^{3}R^{a}{}_{bcd}$, through the $3$-dimensional Ricci identity,
\begin{equation}\label{eq:def_3_riemann}
\, ^{3}R^{a}{}_{bcd}v^{b} =   \left(D_{c}D_{d}-D_{d}D_{c}\right)v^{a} -2 \omega_{cd}u^{p}h_{q}{}^{a}\nabla_{p}v^{q},
\end{equation}
where we used the fact that in a torsion-free spacetime, the $3$-dimensional spatial torsion tensor is proportional to the vorticity $\omega_{ab}$ (see Appendix \ref{appendix_torsion_vorticity} or \cite{Roy:2014lda} for details). By also defining the extrinsic curvature tensor $K_{ab}$,
\begin{equation}\label{eq:def_extrinsic_curvature}
    K_{ab} = D_{a}u_{b}, \qquad K_{c}{}^{a}= g^{ab}K_{cb},
\end{equation}
we can recast the Eq. \eqref{eq:def_3_riemann} as the well known ``Gauss relation'':
\begin{equation}\label{eq:gauss_relation}
\begin{split}
     \, ^{3}R^{a}{}_{bcd} =& h_{c}{}^{m}h_{d}{}^{n}h^{a}{}_{s}h_{b}{}^{r}R^{s}{}_{rmn} + K_{d}{}^{a}K_{cb} +\\
     &- K_{c}{}^{a}K_{db} +2 h_{[c}{}^{m}K_{d]b}h^{an}Q_{mns}u^{s},
\end{split}
\end{equation}
where we have correction terms due to the nonmetricity. The Levi-Civita counterpart of the Gauss relation is instead equal to
\begin{equation}\label{eq:levicivita_gauss_relation}
    \, ^{3}\tilde{R}^{a}{}_{bcd} = h_{c}{}^{m}h_{d}{}^{n}h^{a}{}_{s}h_{b}{}^{r}\tilde{R}^{s}{}_{rmn} + \tilde{K}_{d}{}^{a}\tilde{K}_{cb} - \tilde{K}_{c}{}^{a}\tilde{K}_{db},
\end{equation}
where
\begin{equation}\label{eq:levicivita_def_extrinsic_curvature}
    \tilde{K}_{ab} = \tilde{D}_{a}u_{b}
\end{equation}
is the Levi-Civita extrinsic curvature tensor. The tensors $\, ^{3}R^{a}{}_{bcd}$ and $\, ^{3}\tilde{R}^{a}{}_{bcd}$ are related by the equation 
\begin{equation}\label{eq:gauss_relation_levi_civita_and_total}
\begin{split}
    \, ^{3}R^{a}{}_{bcd} =& \, ^{3}\tilde{R}^{a}{}_{bcd} + 2 \tilde{D}_{[c}L_{d]b}{}^{a} + 2 \tilde{K}_{[c}{}^{a}h_{d]}{}^{m}h_{b}{}^{n}L_{mn}{}^{k}u_{k} +\\
    &+ 2 \tilde{K}_{[c|b}h_{|d]}{}^{m}h^{a}{}_{n}L_{mk}{}^{n}u^{k} +\\
    &+ 2 h_{[c}{}^{r}h_{d]}{}^{m}h_{s}{}^{a}h_{b}{}^{n}h_{l}{}^{k}L_{rk}{}^{s}L_{mn}{}^{l}.
\end{split}    
\end{equation}
Contracting the first and third index in Eq. \eqref{eq:levicivita_gauss_relation}, we have the contracted Levi-Civita Gauss relation:
\begin{equation}\label{eq:levicivita_contracted_gauss_relation}
    \, ^{3}\tilde{R}_{bd} = h_{d}{}^{n}h^{m}{}_{s}h_{b}{}^{r}\tilde{R}^{s}{}_{rmn} + \tilde{K}_{d}{}^{a}\tilde{K}_{ab} - \tilde{K}_{a}{}^{a}\tilde{K}_{db},
\end{equation}
which is useful in deriving the Friedmann equation in homogeneous spacetimes.

\subsubsection{Energy-momentum tensor}

Finally, we remind the decomposition of the (symmetric) energy-momentum tensor $\Psi_{ab}$ with respect to $u^{a}$, $e^{a}$ and $N_{ab}$:
\begin{equation}
\begin{split}
    \Psi_{ab} =& \rho u_{a}u_{b} + 2 q e_{(a}u_{b)} + \left(p-\frac{1}{2}\pi\right) N_{ab} +\\
    &+\left( p + \pi \right) e_{a}e_{b} + 2 \bar{q}_{(a}u_{b)} +  2 \pi_{(a}e_{b)} + \pi_{ab},
\end{split}
\end{equation}
with
\begin{equation}
\begin{gathered}
    \rho = u^{a}u^{b}\Psi_{ab}, \quad q = -e^{a}u^{b}\Psi_{ab}, \quad p = \frac{1}{3}h^{ab}\Psi_{ab}, \\ \pi = \frac{1}{3}\left(2e^{a}e^{b} - N^{ab} \right) \Psi_{ab},\\
    q_{a} = - N_{a}{}^{b}u^{c}\Psi_{bc}, \quad \quad \pi_{a} = N_{a}{}^{b}e^{c} \Psi_{bc}, \\ \pi_{ab} = \left(N_{a}{}^{c}N_{b}{}^{d} - \frac{1}{2}N_{ab}N^{cd}\right)\Psi_{cd} .
\end{gathered}
\end{equation}
The quantities 
\begin{equation}
    p_{\perp} = \left( p - \frac{1}{2} \pi \right) \qquad \text{and} \qquad p_{r} = \left(p + \pi\right)
\end{equation}
represent the transverse and radial pressure of the fluid, respectively.


\section{Locally Rotationally Symmetric spacetime}\label{sec:LRS_spacetime}
A spacetime is said to be Locally Rotationally Symmetric in a neighborhood $N(p^{*})$ of a point $p^{*}$, \emph{if at each point $p$ in $N(p^{*})$ there exists a non-discrete subgroup G of the Lorentz group in the tangent space $T_{p}$ which leaves invariant $u^{a}$, the curvature tensor and their derivatives up to third order} \cite{Stewart:1967tz}. If $G$ is one-dimensional, there is a single preferred spatial direction that constitutes a local axis of symmetry. Therefore, the LRS definition implies that all the spacelike projections of tensors in spatial directions different from the axis of symmetry must be zero.

In the $1+1+2$ formalism, the natural choice for the local axis of symmetry is the above-defined spatial vector $e^{a}$. Hence, the non-zero kinematic and thermodynamic quantities are the scalars
\begin{equation}\label{LRS_scalars}
\begin{gathered}
    \{\theta, \Sigma, \Omega, \phi, \xi, \mathcal{A}^{(u)}, \mathcal{A}^{(e)}, \rho, p, q, \pi,\\
    \mathcal{H}, \check{\bar{\mathcal{H}}}, \bar{\mathcal{H}}, \check{\mathcal{H}}, \mathfrak{H}, \bar{\mathbb{H}}, \check{\mathbb{H}}, \mathcal{E}, \check{\bar{\mathcal{E}}}, \bar{\mathcal{E}}, \check{\mathcal{E}}, \mathfrak{E}, 
    \mathbb{E}, \bar{\mathbb{E}} \}.
\end{gathered}
\end{equation}
In our analysis, we restrict ourselves to an LRS spacetime of type II, so we must require the condition that both the timelike and spacelike congruences are hypersurface orthogonal\footnote{If only the spacelike congruence is hypersurface orthogonal, the spacetime is called LRS of type I. Instead, when only the timelike congruence is hypersurface orthogonal, the corresponding spacetime is LRS of type III.}. In order to satisfy this condition, we assume
\begin{equation}\label{eq:hyper_ortho_u_e}
    \Omega = \xi = 0,
\end{equation}
which, due to the Frobenius theorem, ensures orthogonality (see, for example, \cite{poisson_2004}).

\subsection{Decomposition of nonmetricity tensor in an LRS spacetime of type II}

In an LRS spacetime, the decomposition of the nonmetricity tensor involves only scalar quantities, according to\footnote{In LRS spacetime, both the vector and tensor components are zero.}
\begin{equation}\label{eq:def_nonmetricity_decomposition}
\begin{split}
    Q_{abc} =& -Q_{0}u_{a}u_{b}u_{c} + Q_{1}e_{a}e_{b}e_{c} - Q_{2}u_{a}e_{b}e_{c} +\\
    &- 2 Q_{3} e_{a}u_{(b}e_{c)} + Q_{4}e_{a}u_{b}u_{c} + 2 Q_{5}u_{a}e_{(b}u_{c)} +\\
    &-\frac{1}{2}Q_{6}u_{a}N_{bc} - Q_{7}u_{(b}N_{c)a} + \frac{1}{2}Q_{8}e_{a}N_{bc} +\\
    &+ Q_{9} e_{(b}N_{c)a} - Q_{10} \epsilon_{a(b}u_{c)} + Q_{11}\epsilon_{a(b}e_{c)},
\end{split}
\end{equation}
where
\begin{equation}
\begin{split}
    &Q_{0} = Q_{abc}u^{a}u^{b}u^{c}, \qquad Q_{1} = Q_{abc}e^{a}e^{b}e^{c}, \\ &Q_{2} = Q_{abc}u^{a}e^{b}e^{c}, \qquad
    Q_{3} = Q_{abc}e^{a}e^{b}u^{c}, \\ &Q_{4} = Q_{abc}e^{a}u^{b}u^{c}, \qquad Q_{5} = Q_{abc}u^{a}u^{b}e^{c},\\
    &Q_{6} = Q_{abc}u^{a}N^{bc}, \qquad Q_{7} = Q_{abc}N^{ab}u^{c}, \\  &Q_{8} = Q_{abc}e^{a}N^{bc}, \qquad
    Q_{9} = Q_{abc}N^{ab}e^{c}, \\ &Q_{10} = Q_{abc}\epsilon^{ab}u^{c}, \qquad Q_{11} = Q_{abc}\epsilon^{ab}e^{c}.
\end{split}
\end{equation} 
As a consequence, the nonmetricity scalar \eqref{eq:def_nonmetricity_scalar} is seen to assume the form
\begin{equation}\label{eq:dec_nonmetricity_scalar}
\begin{split}
    \mathcal{Q} =& \frac{1}{2} Q_{0} \left(Q_{3}+Q_{7}\right) + \frac{1}{2}Q_{1}\left( -Q_{5} + Q_{9} \right) +\\
    &+ \frac{1}{2}Q_{2}\left(Q_{3} + Q_{6} - Q_{7} \right) - \frac{1}{2} Q_{4} \left( Q_{5} - Q_{8} + Q_{9} \right) +\\
    &+ \frac{1}{8} Q_{6} \left( -4Q_{3} + Q_{6} \right) - \frac{1}{8} Q_{8} \left(4Q_{5} + Q_{8}\right) +\\
    &- \frac{1}{2}Q_{10}{}^{2} + \frac{1}{2}Q_{11}{}^{2}.
\end{split}
\end{equation}
In view of Eq. \eqref{eq:def_nonmetricity_decomposition}, the Eqs. \eqref{eq:def_covarinat_derivative_u} and \eqref{eq:def_covariant derivative_e} are now written as
\begin{equation}
\begin{split}
    \nabla_{a}u_{b} =& \frac{1}{3}h_{ab}\theta + \left( e_{a}e_{b} - \frac{1}{2}N_{ab} \right)\Sigma - u_{a}e_{b}\mathcal{A}^{(u)} +\\
    &+ \epsilon_{ab} \Omega + \frac{1}{2}u_{a}u_{b}Q_{0} - \frac{1}{2}e_{a}u_{b} Q_{4},
\end{split}
\end{equation}
and
\begin{equation}
\begin{split}
    \nabla_{a}e_{b} =& \frac{1}{2}N_{ab}\phi + \left(\frac{1}{3}\theta + \Sigma\right)e_{a}u_{b} - u_{a}u_{b}\mathcal{A}^{(e)} +\\ 
    &+ \epsilon_{ab}\xi + \frac{1}{2}e_{a}e_{b}Q_{1} - \frac{1}{2}u_{a}e_{b}Q_{2} - e_{a}u_{b}Q_{3}.
\end{split}
\end{equation}
We can also write the commutation relation between the time and spatial derivatives of a generic scalar $\psi$,
\begin{equation}\label{eq:commutation_relation}
\begin{split}
    u^{a}\nabla_{a}\left( \hat{\psi} \right) =& e^{b}\nabla_{b}\left( \dot{\psi} \right) + \dot{\psi} \left( A^{(u)} - \frac{1}{2}Q_{4} \right) +\\
    &- \hat{\psi} \left( \Sigma + \frac{1}{3}\theta + \frac{1}{2}Q_{2} - Q_{3} \right).
\end{split}
\end{equation}
For later use, it is useful to distinguish the contributions due to the Levi-Civita connection from the ones due to the nonmetricity in the following scalars,
\begin{equation}\label{eq:theta_sigma_phi_A_levicivita_nonmetricity}
    \begin{gathered}
    \theta = \tilde{\theta} - \frac{1}{2}Q_{2} + Q_{3} - \frac{1}{2}Q_{6} + Q_{7}, \\ \Sigma = \tilde{\Sigma} - \frac{1}{3}Q_{2} + \frac{2}{3}Q_{3} + \frac{1}{6}Q_{6}- \frac{1}{3}Q_{7}, \\
    \phi = \tilde{\phi} - \frac{1}{2}Q_{8} + Q_{9}, \qquad  \mathcal{A}^{(u)} = \tilde{\mathcal{A}} + \frac{1}{2}Q_{4}, \\  \mathcal{A}^{(e)} = \tilde{\mathcal{A}} + \frac{1}{2}Q_{4} - Q_{5},
    \end{gathered}
\end{equation}
with
\begin{equation}
    \tilde{\theta} = h^{ab}\tilde{D}_{a}u_{b},
\end{equation}
\begin{equation}
    \tilde{\Sigma} = \left[h_{(a}{}^{c}h_{b)}{}^{d} - \frac{1}{3}h_{ab}h^{cd}\right]\tilde{D}_{c}u_{d},
\end{equation}
\begin{equation}
    \tilde{\phi} = N^{ab}\tilde{\delta}_{a}e_{b}. \qquad \tilde{A} = e^{a}u^{b}\tilde{\nabla}_{b}u_{a}.
\end{equation}


\section{\texorpdfstring{$f(\mathcal{Q})$}{} gravity}\label{sec:f_Q_gravity}

The $f(\mathcal{Q})$ gravity is a generalization of the Symmetric Teleparallel Gravity in which both curvature and torsion are zero. The action is given by
\begin{equation}\label{eq:f(Q)_action}
\begin{split}
    A =& \int \hbox{d}^4x \left[-\frac{1}{2} \sqrt{-g} f(\mathcal{Q}) + \lambda_{a}{}^{bcd}R^{a}{}_{bcd} +\right.\\
    &\left. +\lambda_{a}{}^{cd}T_{cd}{}^{a} + \sqrt{-g}\mathcal{L}_{m}\right],
\end{split}
\end{equation}
where $g$ is the determinant of the metric, $f(\mathcal{Q})$ is a generic function of the nonmetricity scalar, $\mathcal{L}_{m}$ is the matter Lagrangian density, $\lambda_{a}{}^{bij}$ and $\lambda_{a}{}^{ij}$ are Lagrange multipliers introduced to impose the vanishing of curvature and torsion. 
Variations with respect to the Lagrange multipliers, metric, and connection yield the field equations,
\begin{equation}\label{eq:lagrange_field_equation}
    R^{a}{}_{bcd} = 0, \qquad  T_{ab}{}^{c} = 0,
\end{equation}
\begin{equation}\label{eq:metric_field_equation}
\begin{split}
    &\frac{2}{\sqrt{-g}}\nabla_{c} \left( \sqrt{-g} f' P^{c}{}_{ab} \right) + \frac{1}{2}g_{ab}f +\\
    & \qquad \qquad + f' \left( P_{acd}Q_{b}{}^{cd} - 2 Q^{cd}{}_{a}P_{cdb} \right) = \Psi_{ab}, 
\end{split}
\end{equation}
and
\begin{equation}\label{eq:connection_field_equation}
    \nabla_{d}\lambda_{c}{}^{bad} +  \lambda_{c}{}^{ab} - \sqrt{-g}f'P^{ab}{}_{c} = \Delta^{ab}{}_{c},
\end{equation}
with 
\begin{equation}
    P^{c}{}_{ab} = -\frac{1}{4}Q^{c}{}_{ab} + \frac{1}{2}Q_{(ab)}{}^{c} + \frac{1}{4}q^{c}g_{ab} - \frac{1}{4}Q^{c}g_{ab} - \frac{1}{4}q_{(a}\delta^{c}_{b)},
\end{equation}
and
\begin{equation}
\begin{gathered}
    \Psi_{ab}=  -\frac{2}{\sqrt{-g}}\frac{\delta \left(\sqrt{-g} \mathcal{L}_{m}\right)}{\delta g^{ab}}, \\
    \Delta^{ab}{}_{c} = - \frac{1}{2}\frac{\delta \left( \sqrt{-g}\mathcal{L}_{m}\right)}{\delta {\Gamma_{ab}{}^{c}}}.
\end{gathered}
\end{equation}
In the above equations, the prime denotes the derivative with respect to the argument of the function. Since we will consider fluids that are independent of nonmetricity, in our case, $\Delta^{ab}{}_{c}$ is identically zero, and the energy-momentum conservation law is equal to the standard one,
\begin{equation}\label{eq:energy-momentum_conservation}
\tilde{\nabla}_{a}\Psi^{a}{}_{b} = 0.
\end{equation}
Because of the flatness and torsionless conditions \eqref{eq:lagrange_field_equation}, from the Eq. \eqref{eq:connection_field_equation} we can also derive the following relation
\begin{equation}\label{eq:connection_field_equations_2}
    \nabla_{a}\nabla_{b}\left(\sqrt{-g} f' P^{ab}{}_{c} + \Delta^{ab}{}_{c}\right)=0,
\end{equation}
which holds on shell. It can be proved that Eq. \eqref{eq:connection_field_equations_2} can be derived by using Eq. \eqref{eq:energy-momentum_conservation} and the Levi-Civita divergence of Eq. \eqref{eq:metric_field_equation}. Hence, for any solution of the metric field equations, Eq. \eqref{eq:connection_field_equations_2} is necessarily satisfied if the energy-momentum conservation holds.

The metric field equation \eqref{eq:metric_field_equation} can be recast in a more suitable form for the $1+1+2$ formalism,
\begin{equation}\label{eq:final_field_equation}
\begin{split}
    \tilde{R}_{ab} =& \frac{1}{f'} \left( \Psi_{ab} - \frac{1}{2}g_{ab}\Psi \right) + \frac{1}{2}g_{ab}\left(\frac{f}{f'} - \mathcal{Q}\right) +\\
    &- 2 \frac{f''}{f'}\left( P^{c}{}_{ab}-\frac{1}{2}g_{ab}P^{cd}{}_{d} \right)\partial_{c}\mathcal{Q},
\end{split}
\end{equation}
with $\Psi=g^{ab}\Psi_{ab}$. By replacing $f(\mathcal{Q}) = \mathcal{Q}$ into Eq. \eqref{eq:final_field_equation}, we recover the field equations of GR,
\begin{equation}
    \tilde{R}_{ab} = \Psi_{ab} - \frac{1}{2}g_{ab}\Psi.
\end{equation}


\subsection{\texorpdfstring{$1+1+2$}{} field equations in LRS spacetimes of type II}
To make explicit the field equations in the $1+1+2$ formalism, we need the Ricci identity for the Levi-Civita Riemann tensor, written for both $u_{a}$ and $e_{a}$,
\begin{equation}\label{eq:def_ricci_identity}
\begin{split}
    \left[\tilde{\nabla}_{c}\tilde{\nabla}_{d} - \tilde{\nabla}_{d}\tilde{\nabla}_{c}\right]u_{b} =& - \tilde{R}^{a}{}_{bcd}u_{a}, \\ \left[\tilde{\nabla}_{c}\tilde{\nabla}_{d} - \tilde{\nabla}_{d}\tilde{\nabla}_{c}\right]e_{b} =& - \tilde{R}^{a}{}_{bcd}e_{a} .
\end{split}
\end{equation}
From the Eqs. \eqref{eq:final_field_equation} and \eqref{eq:def_ricci_identity}, we obtain the following relations:
\begin{itemize}
    \item $\tilde{R}_{ad}u^{a}u^{d} = g^{bc}u^{d} \left[ \tilde{\nabla}_{c}\tilde{\nabla}_{d} - \tilde{\nabla}_{d}\tilde{\nabla}_{c} \right] u_{b}$,
    \begin{equation}\label{eq:levicivita_ricci_uu}
    \begin{split}
    &\hat{\tilde{\mathcal{A}}} - \dot{\tilde{\theta}} + \tilde{\mathcal{A}}^{2}  - \frac{1}{3}\tilde{\theta}^{2} - \frac{3}{2}\tilde{\Sigma}^{2} + \tilde{\mathcal{A}}\tilde{\phi}- \frac{1}{2}\frac{1}{f'}\left( \rho + 3p \right) + \frac{1}{2}\frac{f}{f'} +\\
    &- \frac{1}{2}\mathcal{Q} - \frac{1}{2}\frac{f''}{f'}\left(Q_{4} - 2Q_{5} \right)\hat{\mathcal{Q}} - \frac{1}{2}\frac{f''}{f'}\left(Q_{2} + Q_{6} \right)\dot{\mathcal{Q}} = 0 ;
    \end{split}    
    \end{equation}
    \item $\tilde{R}_{ad}e^{a}e^{d} = g^{bc}e^{d} \left[ \tilde{\nabla}_{c}\tilde{\nabla}_{d} - \tilde{\nabla}_{d}\tilde{\nabla}_{c} \right] e_{b}$,
    \begin{equation}\label{eq:levicivita_ricci_ee}
    \begin{split}
        &\hat{\tilde{\phi}} + \hat{\tilde{A}} - \frac{1}{3}\dot{\tilde{\theta}} - \dot{\tilde{\Sigma}} + \frac{1}{2}\tilde{\phi}^{2}  + \tilde{A}^{2} - \frac{1}{3}\tilde{\theta}^{2} - \tilde{\theta}\tilde{\Sigma} +\\
        &+ \frac{1}{2}\frac{1}{f'}\left( \rho - p + 2 \pi \right)  +\frac{1}{2}\frac{f}{f'} - \frac{1}{2}\mathcal{Q} +\\
        &- \frac{1}{2}\frac{f''}{f'}\left(Q_{4} - Q_{8} \right)\hat{\mathcal{Q}} - \frac{1}{2}\frac{f''}{f'}\left(Q_{2} - 2Q_{3} \right)\dot{\mathcal{Q}} = 0 ;
    \end{split}
    \end{equation}
    \item $\tilde{R}_{ad}u^{a}e^{d} = g^{bc}e^{d} \left[ \tilde{\nabla}_{c}\tilde{\nabla}_{d} - \tilde{\nabla}_{d}\tilde{\nabla}_{c} \right] u_{b}$,
    \begin{equation}\label{eq:levicivita_ricci_ue}
    \begin{split}
    &\hat{\tilde{\theta}} - \frac{3}{2}\hat{\tilde{\Sigma}} - \frac{9}{4}\tilde{\Sigma}\tilde{\phi} - \frac{3}{8} \frac{f''}{f'}\left(Q_{0} + Q_{2} - Q_{6}\right)\hat{\mathcal{Q}} +\\
    &+ \frac{3}{8} \frac{f''}{f'}\left(Q_{1} + Q_{4} + Q_{8}\right)\dot{\mathcal{Q}} = \frac{3}{2} \frac{1}{f'} q; 
    \end{split}
    \end{equation}
    \indent
    \item $\tilde{R}_{ad}e^{a}u^{d} = g^{bc}u^{d} \left[ \tilde{\nabla}_{c}\tilde{\nabla}_{d} - \tilde{\nabla}_{d}\tilde{\nabla}_{c} \right] e_{b}$,
    \begin{equation}\label{eq:levicivita_ricci_eu}
    \begin{split}
    &\dot{\tilde{\phi}} + \frac{1}{3} \tilde{\phi} \left( \tilde{\theta} - \frac{3}{2} \tilde{\Sigma}  \right) + \tilde{A} \left( \tilde{\Sigma} - \frac{2}{3}\tilde{\theta} \right) +\\  
    &- \frac{3}{4} \frac{f''}{f'}\left(Q_{0} + Q_{2} - Q_{6}\right)\hat{\mathcal{Q}} +\\
    &+ \frac{3}{4} \frac{f''}{f'}\left(Q_{1} + Q_{4} + Q_{8}\right)\dot{\mathcal{Q}} = \frac{1}{f'}q; 
    \end{split}
    \end{equation}
    \item $-\tilde{R}^{a}{}_{bcd}e_{a}u^{b}u^{c}e^{d} = u^{b}u^{c}e^{d} \left[ \tilde{\nabla}_{c}\tilde{\nabla}_{d} - \tilde{\nabla}_{d}\tilde{\nabla}_{c} \right] e_{b}$,
    \begin{equation}\label{eq:levicivita_riemann_contraction}
    \begin{split}
    &\hat{\tilde{A}} - \frac{1}{3}\dot{\tilde{\theta}} - \dot{\tilde{\Sigma}} + \tilde{A}^{2} - \frac{1}{9}\tilde{\theta}^{2} - \tilde{\Sigma}^{2} - \frac{2}{3}\tilde{\theta}\tilde{\Sigma}  +\\
    &- \tilde{\mathcal{E}} -\frac{1}{6}\frac{1}{f'}\left( \rho + 3 p - 3 \pi \right) + \frac{1}{6}\frac{f}{f'} - \frac{1}{6}\mathcal{Q} +\\
    &- \frac{1}{12}\frac{f''}{f'}\left(4 Q_{4} - 4 Q_{5} - Q_{8} - 2 Q_{9}\right)\hat{Q} +\\
    &- \frac{1}{12}\frac{f''}{f'}\left(4 Q_{2} - 4 Q_{3} + Q_{6} + 2 Q_{7}\right)\dot{Q} = 0;
    \end{split}
    \end{equation}
    \indent
    \item $-\tilde{R}^{a}{}_{bcd}e_{a}u^{b}\epsilon^{cd} = \epsilon^{cd}u^{b} \left[ \tilde{\nabla}_{c}\tilde{\nabla}_{d} - \tilde{\nabla}_{d}\tilde{\nabla}_{c} \right] e_{b}$,
    \begin{equation}\label{eq:levicivita_riemann_contraction_epsilon}
    \tilde{\mathcal{H}} = 0.\\
    \end{equation}
\end{itemize}
\indent
\subsection{Weyl tensor}
It is helpful to make explicit the contributions due to the Levi-Civita connection and the nonmetricity tensor in the magnetic and electric parts of the Weyl tensor. In this way,
keeping in mind that the Weyl tensor is zero because of the Eq. \eqref{eq:lagrange_field_equation}, we can derive constraints on the nonmetricity from the scalar parts of the Eqs. \eqref{eq:dec_magnetic_weyl_1}-\eqref{eq:dec_electric_weyl_3}:
\begin{widetext}
\begin{equation}\label{eq:magnetic_part_weyl_1}
\begin{split}
    \mathcal{H} = 0 =& \tilde{\mathcal{H}} + \frac{1}{4}\hat{Q}_{10} - \frac{1}{4}\dot{Q}_{11} - Q_{10}\left( \frac{\tilde{\phi}}{8} - \frac{\tilde{A}}{4} + \frac{Q_{8}}{16} + \frac{Q_{9}}{8} \right) - Q_{11} \left( \frac{3\tilde{\Sigma}}{8} - \frac{Q_{6}}{16} - \frac{Q_{7}}{8} \right);
\end{split}
\end{equation}
\begin{equation}\label{eq:magnetic_part_weyl_2}
\begin{split}
    \check{\bar{\mathcal{H}}} = \mathfrak{H} = 0 =& \frac{7}{24} \dot{Q}_{10} - \frac{1}{24}\hat{Q}_{11} + Q_{10} \left( \frac{\tilde{\Sigma}}{16} - \frac{7Q_{0}}{48} - \frac{Q_{2}}{48} + \frac{Q_{3}}{24} + \frac{11Q_{6}}{96} - \frac{11Q_{7}}{48} \right) +\\
    &- Q_{11} \left( \frac{5\tilde{\phi}}{48} + \frac{7\tilde{A}}{24} + \frac{Q_{1}}{48} + \frac{7Q_{4}}{24} - \frac{7Q_{5}}{24} - \frac{7Q_{8}}{96} + \frac{7Q_{9}}{48} \right); 
\end{split}
\end{equation}
\begin{equation}\label{eq:magnetic_part_weyl_3}
\begin{split}
    \bar{\mathcal{H}} = 0 =& \tilde{\mathcal{H}} + \frac{3}{8}\hat{Q}_{10} - \frac{1}{8}\dot{Q}_{11} - Q_{10} \left( \frac{3\tilde{\phi}}{16} - \frac{\tilde{A}}{8} + \frac{Q_{4}}{8} - \frac{Q_{8}}{32} + \frac{5 Q_{9}}{16} \right) - Q_{11} \left( \frac{\tilde{\theta}}{12} + \frac{7\tilde{\Sigma}}{16} - \frac{Q_{2}}{8} - \frac{5Q_{6}}{32} + \frac{Q_{7}}{16} \right);
\end{split}
\end{equation}
\begin{equation}\label{eq:magnetic_part_weyl_4}
\begin{split}
    \check{\mathcal{H}} = 0 =& -\tilde{\mathcal{H}} - \frac{1}{8}\hat{Q}_{10} + \frac{3}{8}\hat{Q}_{11} + Q_{10} \left( \frac{\tilde{\phi}}{16} - \frac{3\tilde{A}}{8} - \frac{Q_{4}}{8} + \frac{5Q_{8}}{32} - \frac{Q_{9}}{16} \right) - Q_{11} \left( \frac{\tilde{\theta}}{12} - \frac{5\tilde{\Sigma}}{16} - \frac{Q_{2}}{8} - \frac{Q_{6}}{32} + \frac{5Q_{7}}{16} \right);
\end{split}
\end{equation}
\begin{equation}\label{eq:magnetic_part_weyl_5}
\begin{split}
    \bar{\mathbb{H}} = \check{\mathbb{H}} = 0 =& -\frac{5}{24}\hat{Q}_{0} + \frac{1}{24}\dot{Q}_{1} - \frac{1}{24}\hat{Q}_{2} + \frac{5}{24}\dot{Q}_{4} - \frac{1}{12}\hat{Q}_{6} - \frac{5}{24}\hat{Q}_{7} + \frac{1}{12}\dot{Q}_{8} - \frac{1}{24}\dot{Q}_{9} + \tilde{\phi}\left( \frac{Q_{2}}{24} + \frac{5Q_{3}}{24} - \frac{Q_{6}}{48} - \frac{5Q_{7}}{48} \right) +\\
    &- \tilde{A} \left( \frac{5Q_{0}}{24} + \frac{Q_{2}}{24} + \frac{Q_{3}}{2} + \frac{Q_{6}}{12} - \frac{Q_{7}}{24} \right) + \tilde{\theta} \left( \frac{Q_{1}}{72} - \frac{5Q_{4}}{72} + \frac{5Q_{5}}{36} - \frac{Q_{8}}{24} + \frac{Q_{9}}{18} \right) +\\
    &+ \tilde{\Sigma} \left( \frac{Q_{1}}{24} + \frac{5Q_{4}}{12} + \frac{13Q_{5}}{24} + \frac{3Q_{8}}{16} + \frac{11Q_{9}}{48} \right) - Q_{2} \left( \frac{Q_{5}}{4} + \frac{Q_{8}}{48} + \frac{Q_{9}}{12} \right) - Q_{3}\left( \frac{Q_{4}}{4} - \frac{Q_{5}}{2} + \frac{5Q_{8}}{48} - \frac{5Q_{9}}{24} \right) +\\
    &+ Q_{4} \left( \frac{5Q_{6}}{48} - \frac{Q_{7}}{12} \right) + Q_{5} \left( \frac{Q_{6}}{48} - \frac{Q_{7}}{24} \right) + Q_{8} \left( \frac{Q_{6}}{16} - \frac{5Q_{7}}{96} \right) + \frac{1}{8}Q_{10}Q_{11};
\end{split}
\end{equation}
\begin{equation}\label{eq:electric_part_weyl_1}
\begin{split}
    \mathcal{E} = 0 =& \tilde{\mathcal{E}} + \frac{1}{6}\dot{Q}_{2} + \frac{1}{12}\dot{Q}_{3} + \frac{1}{6}\hat{Q}_{4} - \frac{5}{12}\hat{Q}_{5} - \frac{1}{12}\dot{Q}_{6} - \frac{1}{24}\dot{Q}_{7} + \frac{1}{12}\hat{Q}_{8} - \frac{5}{24}\hat{Q}_{9} - \frac{Q_{2}^{2}}{8} - \frac{Q_{4}^{2}}{8} + \frac{Q_{6}^{2}}{96} + \frac{5Q_{8}^{2}}{96} + \frac{5Q_{10}^{2}}{48} +\\
    &+ \frac{Q_{11}^{2}}{48} + \tilde{\phi}\left( \frac{Q_{1}}{8} - \frac{Q_{4}}{12} + \frac{5Q_{5}}{24} - \frac{5Q_{8}}{48} - \frac{Q_{9}}{48} \right) - \tilde{A} \left( \frac{Q_{1}}{4} + \frac{Q_{4}}{12} + \frac{5Q_{5}}{12} - \frac{Q_{8}}{12} - \frac{Q_{9}}{24}  \right) + \tilde{\theta} \left( \frac{Q_{2}}{12} - \frac{Q_{6}}{24} \right) +\\
    &+ \tilde{\Sigma} \left(\frac{3Q_{0}}{8} + \frac{Q_{2}}{2} + \frac{Q_{3}}{8} - \frac{Q_{6}}{16} + \frac{5Q_{7}}{16} \right) - Q_{0} \left( \frac{Q_{2}}{8} - \frac{Q_{3}}{24} - \frac{Q_{6}}{16} + \frac{Q_{7}}{48} \right) - Q_{1}\left( \frac{Q_{4}}{8} - \frac{5Q_{5}}{24} + \frac{Q_{8}}{16} - \frac{5Q_{9}}{48} \right) +\\
    &+ Q_{2} \left( \frac{Q_{3}}{24} + \frac{Q_{6}}{24} - \frac{5Q_{7}}{48} \right) + Q_{4} \left( \frac{5Q_{5}}{24} + \frac{Q_{8}}{24} + \frac{Q_{9}}{48} \right) + Q_{6} \left( \frac{Q_{3}}{48} + \frac{Q_{7}}{32} \right) - Q_{8} \left( \frac{5Q_{5}}{48} + \frac{3Q_{9}}{32} \right);
\end{split}    
\end{equation}
\begin{equation}\label{eq:electric_part_weyl_2}
\begin{split}
    \bar{\mathcal{E}} = 0 =& -\tilde{\mathcal{E}} - \frac{1}{6}\dot{Q}_{2} + \frac{5}{12}\dot{Q}_{3} - \frac{1}{6}\hat{Q}_{4} - \frac{1}{12}\hat{Q}_{5} + \frac{1}{12}\dot{Q}_{6} - \frac{5}{24}\dot{Q}_{7} - \frac{1}{12}\hat{Q}_{8} - \frac{1}{24}\hat{Q}_{9} - \frac{Q_{2}^{2}}{8} - \frac{Q_{4}^{2}}{8} + \frac{5Q_{6}^{2}}{96} + \frac{Q_{8}^{2}}{96} + \frac{Q_{10}^{2}}{48} +\\ 
    &+ \frac{5Q_{11}^{2}}{48} + \tilde{\phi}\left( \frac{Q_{1}}{8} + \frac{Q_{4}}{12} + \frac{Q_{5}}{24} - \frac{Q_{8}}{48} - \frac{5Q_{9}}{48} \right) - \tilde{A} \left( \frac{Q_{1}}{4} + \frac{5Q_{4}}{12} + \frac{Q_{5}}{12} + \frac{Q_{8}}{12} - \frac{5Q_{9}}{24}  \right) + \tilde{\theta} \left( \frac{Q_{2}}{12} - \frac{Q_{6}}{24} \right) +\\
    &+ \tilde{\Sigma} \left(\frac{3Q_{0}}{8} + \frac{5Q_{3}}{8} + \frac{3Q_{6}}{16} + \frac{Q_{7}}{16} \right) - Q_{0} \left( \frac{Q_{2}}{8} - \frac{5Q_{3}}{24} - \frac{Q_{6}}{16} + \frac{5Q_{7}}{48} \right) - Q_{1}\left( \frac{Q_{4}}{8} - \frac{Q_{5}}{24} + \frac{Q_{8}}{16} - \frac{Q_{9}}{48} \right) +\\
    &+ Q_{2} \left( \frac{5Q_{3}}{24} - \frac{Q_{6}}{24} - \frac{Q_{7}}{48} \right) + Q_{4} \left( \frac{Q_{5}}{24} - \frac{Q_{8}}{24} + \frac{5Q_{9}}{48} \right) + Q_{6} \left( \frac{5Q_{3}}{48} - \frac{3Q_{7}}{32} \right) - Q_{8} \left( \frac{Q_{5}}{48} - \frac{Q_{9}}{32} \right);
\end{split}
\end{equation}
\begin{equation}\label{eq:electric_part_weyl_3}
\begin{split}
    \check{\mathcal{E}} = 0 =& \frac{5}{24}\hat{Q}_{0} - \frac{1}{24}\dot{Q}_{1} + \frac{1}{24}\hat{Q}_{2} - \frac{5}{24}\dot{Q}_{4} + \frac{1}{12}\hat{Q}_{6} + \frac{5}{24}\hat{Q}_{7} - \frac{1}{12}\dot{Q}_{8} + \frac{1}{24}\dot{Q}_{9} +\\
    &-\tilde{\phi}\left( \frac{Q_{2}}{24} + \frac{5Q_{3}}{24} - \frac{Q_{6}}{48} - \frac{5Q_{7}}{48} \right) + \tilde{A} \left( \frac{5Q_{0}}{24} + \frac{Q_{2}}{24} + \frac{Q_{3}}{2} + \frac{Q_{6}}{12} - \frac{Q_{7}}{24}  \right) - \tilde{\theta} \left( \frac{Q_{1}}{72} - \frac{5Q_{4}}{72} + \frac{5Q_{5}}{36} - \frac{Q_{8}}{24} + \frac{Q_{9}}{18} \right) +\\
    &- \tilde{\Sigma} \left(\frac{Q_{1}}{24} + \frac{5Q_{4}}{12} + \frac{13Q_{5}}{24} + \frac{3Q_{8}}{16} + \frac{11Q_{9}}{48} \right) - Q_{0} \left( \frac{Q_{2}}{8} - \frac{5Q_{3}}{24} - \frac{Q_{6}}{16} + \frac{5Q_{7}}{48} \right) - Q_{1}\left( \frac{Q_{4}}{8} - \frac{Q_{5}}{24} + \frac{Q_{8}}{16} - \frac{Q_{9}}{48} \right) +\\
    &+ Q_{2} \left( \frac{5Q_{3}}{24} - \frac{Q_{6}}{24} - \frac{Q_{7}}{48} \right) + Q_{4} \left( \frac{Q_{5}}{24} - \frac{Q_{8}}{24} + \frac{5Q_{9}}{48} \right) + Q_{6} \left( \frac{5Q_{3}}{48} - \frac{3Q_{7}}{32} \right) - Q_{8} \left( \frac{Q_{5}}{48} - \frac{Q_{9}}{32} \right) +\\
    &- Q_{3} \left( \frac{Q_{5}}{2} - \frac{5Q_{8}}{48} + \frac{5Q_{9}}{24}\right) - Q_{5} \left( \frac{Q_{6}}{48} - \frac{Q_{7}}{24} \right) - Q_{6} \left( \frac{Q_{8}}{16} - \frac{Q_{9}}{96} \right) + \frac{5}{96}Q_{7}Q_{8} - \frac{1}{8}Q_{10}Q_{11};
\end{split}
\end{equation}
\begin{equation}\label{eq:electric_part_weyl_4}
\begin{split}
    \mathbb{E} = \bar{\mathbb{E}} = 0 =& -\frac{7}{48}\dot{Q}_{10} + \frac{1}{48}\hat{Q}_{11} - Q_{10}\left( \frac{\tilde{\Sigma}}{32} - \frac{7Q_{0}}{96} - \frac{Q_{2}}{96} + \frac{Q_{3}}{48} + \frac{11Q_{6}}{192} - \frac{11Q_{7}}{96} \right) +\\
    &+ Q_{11} \left( \frac{5\tilde{\phi}}{96} + \frac{7\tilde{A}}{48} + \frac{Q_{1}}{96} + \frac{7Q_{4}}{96} - \frac{7Q_{5}}{48} - \frac{7Q_{8}}{192} + \frac{7Q_{9}}{96} \right);
\end{split}
\end{equation}
\begin{equation}\label{eq:electric_part_weyl_5}
\begin{split}
    \check{\mathbb{E}} = 0 =& \frac{7}{24}\dot{Q}_{10} - \frac{1}{24}\hat{Q}_{11} + Q_{10}\left( \frac{\tilde{\Sigma}}{16} - \frac{7Q_{0}}{48} - \frac{Q_{2}}{48} + \frac{Q_{3}}{24} + \frac{11Q_{6}}{96} - \frac{11Q_{7}}{48} \right) +\\
    &- Q_{11} \left( \frac{5\tilde{\phi}}{48} + \frac{7\tilde{A}}{24} + \frac{Q_{1}}{48} + \frac{7Q_{4}}{48} - \frac{7Q_{5}}{24} - \frac{7Q_{8}}{96} + \frac{7Q_{9}}{48} \right).
\end{split}
\end{equation}

\end{widetext}

\subsection{Riemann tensor contractions}
Besides the Weyl tensor, we can deduce constraints from the Riemann tensor itself by considering all its possible contractions with $u^{a}$, $e^{a}$, $N^{ab}$, and imposing Eq. \eqref{eq:lagrange_field_equation}. Making use of the Ricci identities for the full connection,
\begin{equation}\label{eq:def_ricci_identity_total_connection}
\begin{split}
    \left[\nabla_{c}\nabla_{d} - \nabla_{d}\nabla_{c}\right]u_{b} =& - R^{a}{}_{bcd}u_{a}, \\
    \left[\nabla_{c}\nabla_{d} - \nabla_{d}\nabla_{c}\right]e_{b} =& - R^{a}{}_{bcd}e_{a},
\end{split}
\end{equation}
we deduce the following relations:
\begin{widetext}
\begin{itemize}
    \item $R^{a}{}_{bcd}u_{a}u^{b}u^{c}e^{d}=0$,
    \begin{equation}\label{eq:riemann_contraction_1}
    \hat{Q}_{0} - \dot{Q}_{4} + \mathcal{A}^{(u)} \left( Q_{0} + 2 Q_{3} \right) - \frac{1}{3} \theta  \left( Q_{4} + 2Q_{5} \right) + \Sigma \left( 3\mathcal{A}^{(e)} - 3\mathcal{A}^{(u)} - Q_{4} + Q_{5} \right) - Q_{4} \left( \frac{Q_{0}}{2} + \frac{Q_{2}}{2} + Q_{3} \right) = 0;
    \end{equation}
    \item $R^{a}{}_{bcd}e_{a}e^{b}u^{c}e^{d}=0$,
    \begin{equation}\label{eq:riemann_contraction_2}
    \begin{split}
    \hat{Q}_{2} - \dot{Q}_{1} + \mathcal{A}^{(u)} \left( Q_{2} + 2Q_{3} \right) + 2 \left( \Sigma + \frac{1}{3} \theta\right) \left( \mathcal{A}^{(e)} - \mathcal{A}^{(u)} - \frac{1}{2} Q_{1} \right) - \frac{1}{2} Q_{1} Q_{2} +  Q_{1} Q_{3} - \frac{1}{2} Q_{2} Q_{4} = 0;
    \end{split}
    \end{equation}
    \item $R^{a}{}_{bcd}N_{a}{}^{b}u^{c}e^{d}=0$,
    \begin{equation}\label{eq:riemann_contraction_3}
    \begin{split}
    & \hat{Q}_{6} - \dot{Q}_{8} + \tilde{\mathcal{A}}Q_{6} - \frac{1}{3}\tilde{\theta}Q_{8} - \tilde{\Sigma}Q_{8} = 0;
    \end{split}
    \end{equation}
    \item $R^{a}{}_{bcd}u_{a}e^{b}u^{c}e^{d}=0$,
    \begin{equation}\label{eq:riemann_contraction_4}
    \hat{\mathcal{A}}^{(u)} - \frac{1}{3}\dot{\theta} - \dot{\Sigma} + \mathcal{A}^{(u)}{}^{2} + \frac{1}{2} \mathcal{A}^{(u)} Q_{1} - \left( \Sigma + \frac{1}{3}\theta \right)^{2} - \frac{1}{2}\left(\Sigma + \frac{1}{3}\theta\right) \left( Q_{0} + 2Q_{2} - 2Q_{3} \right) = 0;
    \end{equation}
    \item $R^{a}{}_{bcd}e_{a}u^{b}u^{c}e^{d}=0$,
    \begin{equation}\label{eq:riemann_contraction_5}
    \begin{split}
    &\hat{\mathcal{A}}^{(e)} - \frac{1}{3}\dot{\theta} - \dot{\Sigma} + \dot{Q}_{3} + \mathcal{A}^{(e)} \mathcal{A}^{(u)} - \frac{1}{2} \mathcal{A}^{(e)} \left( Q_{1} + 2 Q_{4} \right) + \frac{1}{2} \phi \left( \mathcal{A}^{(e)} - \mathcal{A}^{(u)} + Q_{5} \right) - \left( \Sigma + \frac{1}{3}\theta \right)^{2} +\\
    &+ \frac{1}{2} \left( \Sigma + \frac{1}{3} \theta \right) \left( Q_{0} + 4 Q_{3} \right) - Q_{3}^{2} - \frac{1}{2} Q_{0} Q_{3} = 0;
    \end{split}
    \end{equation}
    \item $R^{a}{}_{bcd}N_{a}{}^{c}u^{b}u^{d}=0$,
    \begin{equation}\label{eq:riemann_contraction_6}
    \begin{split}
    &\dot{\Sigma} - -\frac{2}{3}\dot{\theta} + \dot{Q}_{7} - \frac{1}{2} \left( \Sigma - \frac{2}{3}\theta \right)^{2} - \frac{1}{2} \left( \Sigma - \frac{2}{3}\theta \right) \left( Q_{0} + 2 Q_{7} \right) + \phi \left( A^{(u)} - Q_{5} \right) +\\
    &+ 2\phi Q_{5} + 2\mathcal{A}^{(e)} Q_{9} - \frac{1}{2} Q_{0} Q_{7} - \frac{1}{2} Q_{7}^{2} + \frac{1}{2}Q_{10}^{2} = 0;
    \end{split}
    \end{equation}
    \item $R^{a}{}_{bcd}N_{a}{}^{c}e^{b}e^{d}=0$,
    \begin{equation}\label{eq:riemann_contraction_7}
    \begin{split}
    \hat{\phi} - \hat{Q}_{9} + \frac{\phi^{2}}{2} + \Sigma^{2} - \frac{2}{9} \theta^{2} - \frac{1}{3} \theta  \Sigma + \frac{1}{2} \phi \left( Q_{1} - 2 Q_{9} \right) + \frac{1}{3} \theta  Q_{7} + \Sigma Q_{7} - \frac{1}{2} Q_{1} Q_{9} + \frac{1}{2} Q_{9}^{2} - \frac{1}{2}Q_{11}^{2} = 0;
    \end{split}
    \end{equation}
    \item $R^{a}{}_{bcd}N_{a}{}^{c}u^{b}e^{d}=0$,
    \begin{equation}\label{eq:riemann_contraction_8}
   \begin{split}
    \hat{\Sigma} - \frac{2}{3} \hat{\theta} + \hat{Q}_{7} + \frac{3}{2} \Sigma \phi + \frac{1}{3} \theta Q_{4} - \frac{1}{2} \Sigma \left( Q_{4} + 3 Q_{9} \right) - \left( \phi - Q_{9} \right) \left( Q_{3} - \frac{1}{2} Q_{7} \right)  - \frac{1}{2} Q_{4} Q_{7} + \frac{1}{2}Q_{10}Q_{11} = 0;
    \end{split}
    \end{equation}
    \item $R^{a}{}_{bcd}N_{a}{}^{c}e^{b}u^{d}=0$,
    \begin{equation}\label{eq:riemann_contraction_9}
    \begin{split}
    \dot{\phi} - \dot{Q}_{9} - \frac{1}{2}\left( \Sigma - \frac{2}{3}\theta \right) \left( \phi - 2\mathcal{A}^{(e)} - 2Q_{5} - Q_{9} \right) + \frac{1}{2} \left( \phi - Q_{9} \right) \left( Q_{2} - Q_{7} \right)  + \mathcal{A}^{(u)} Q_{7} - \frac{1}{2}Q_{10}Q_{11} = 0;
    \end{split}
    \end{equation}
    \item $R^{a}{}_{bcd}N^{bc}u_{a}u^{d}=0$,
    \begin{equation}\label{eq:riemann_contraction_10}
    \begin{split}
    \dot{\Sigma} - \frac{2}{3}\dot{\theta} + \phi \mathcal{A}^{(u)} + -\frac{1}{2}\left(\Sigma - \frac{2}{3}\theta\right)^{2}  + \frac{1}{2} \left(\Sigma - \frac{2}{3} \theta \right) \left( Q_{0} + Q_{6} - Q_{7} \right) = 0;
    \end{split}
    \end{equation}
    \item $R^{a}{}_{bcd}N^{bc}e_{a}e^{d}=0$,
    \begin{equation}\label{eq:riemann_contraction_11}
    \begin{split}
    \hat{\phi} + \frac{\phi ^{2}}{2} + \Sigma^{2} - \frac{2}{9}\theta^{2} - \frac{1}{3} \theta  \Sigma - \frac{1}{2} \phi  \left( Q_{1} - Q_{8} + Q_{9}\right) - \Sigma  Q_{3} + \frac{2}{3} \theta  Q_{3}  = 0;
    \end{split}
    \end{equation}
    \item $R^{a}{}_{bcd}N^{bc}u_{a}e^{d}=0$,
    \begin{equation}\label{eq:riemann_contraction_12}
    \begin{split}
    \hat{\Sigma} - \frac{2}{3} \hat{\theta} + \frac{3}{2} \Sigma  \phi  + \frac{1}{2} \left( \Sigma -\frac{2}{3}\theta \right) \left( Q_{4} + Q_{8} - Q_{9} \right) = 0;
    \end{split}
    \end{equation}
    \item $R^{a}{}_{bcd}N^{bc}e_{a}u^{d}=0$,
    \begin{equation}\label{eq:riemann_contraction_13}
    \begin{split}
    \dot{\phi} - \frac{1}{2}\phi  \left( \Sigma - \frac{2}{3}\theta + Q_{2} - Q_{6} + Q_{7} \right) + \mathcal{A}^{(e)} \left(\Sigma - \frac{2}{3}\theta \right) = 0;
    \end{split}
    \end{equation}
\end{itemize}
\end{widetext}
All the remaining contractions, which are not explicitly expressed above, are identically null.


\section{Homogeneous spacetime}\label{sec:hom_spacetime}
The spacetime is said ``homogeneous'' \cite{Weinberg:1972kfs, Wald:1984r} \emph{if there exists a one-parameter family of spacelike hypersurfaces $\Sigma_{t}$ foliating the spacetime such that for each $t$ and for any points $P$, $Q$ $\in \Sigma_{t}$, there exists an isometry of the metric tensor $g_{ab}$ which takes $P$ into $Q$.}
Therefore, the spacetime must admit a spacelike Killing vector field $\mathbf{X}$,
\begin{equation}\label{eq:hom_lie_derivative_metric_1_1_2}
    \mathcal{L}_{\mathbf{X}} g_{ab} = 0,
\end{equation}
with $\mathcal{L}_{\mathbf{X}}$ denoting Lie derivative. In our geometrical setting in which the torsion tensor is zero, we also require the further condition 
\begin{equation}\label{eq:hom_lie_derivative_nonmetricity_1_1_2}
    \mathcal{L}_{\mathbf{X}} Q_{abc} = 0,
\end{equation}
i.e., both the metric and nonmetricity tensors are invariant under the action of the $1$-parameter group of local diffeomorphisms generated by $\mathbf{X}$.

In the presence of nonmetricity, the Killing equations assume the form
\begin{equation}\label{eq:def_killing_equation_nonmetricity}
\begin{split}
    0 = \mathcal{L}_{\mathbf{X}}g_{ab} =& 2\nabla_{(a}X_{b)} + 2X^{c}L_{abc} =\\ =& 2\tilde{\nabla}_{(a}X_{b)} .
\end{split}
\end{equation}
The form involving the total covariant derivative is the one used in the following analysis.

We choose the Killing vector expressed as
\begin{equation}\label{eq:hom_killing_vect_cond_1}
    \mathbf{X} = C(\mathbf{x})e,
\end{equation}
where $C(\mathbf{x})$ is a generic smooth function of the coordinates. Contracting the Eq. \eqref{eq:def_killing_equation_nonmetricity} twice with $u^{a}$, we obtain 
\begin{equation}\label{eq:hom_cond_Au}
    A^{(u)} = \frac{1}{2}Q_{4}.
\end{equation} 
The contraction of the Eq. \eqref{eq:def_killing_equation_nonmetricity} with $u^{a}$ and $e^{b}$ yields the relation
\begin{equation}\label{eq:hom_killing_vect_cond_2}
    \dot{C} = C\left( \Sigma + \frac{1}{3}\Theta + \frac{1}{2}Q_{2} - Q_{3} \right),
\end{equation}
while the double contraction with the spatial vector $e^{a}$ gives us
\begin{equation}\label{eq:hom_killing_vect_cond_3}
    \hat{C} = 0.
\end{equation}
Finally, from the contraction with $N^{ab}$ we find
\begin{equation}\label{eq:hom_cond_Phi}
    \phi = -\frac{1}{2}Q_{8} + Q_{9}.
\end{equation}
On the other hand, the Lie derivative of the nonmetricity \eqref{eq:hom_lie_derivative_nonmetricity_1_1_2} ensures that the spatial derivative of the scalar components of the nonmetricity tensor is zero,
\begin{equation}
    \hat{Q}_{i} = 0, \qquad i=0,...,11.
\end{equation}
Hence, because of the LRS symmetries and the homogeneity condition, the covariant derivatives of the scalars \eqref{LRS_scalars} are zero except for those along the temporal direction $u^{a}$.

It is worth mentioning that Eq. \eqref{eq:hom_cond_Au} can also be derived by the commutation relation \eqref{eq:commutation_relation} by imposing that the hat derivative is null.

By directly comparing Eq. \eqref{eq:theta_sigma_phi_A_levicivita_nonmetricity} with the Eqs. \eqref{eq:hom_cond_Au} and \eqref{eq:hom_cond_Phi}, we obtain the following conditions on $\tilde{A}$ and $\tilde{\phi}$:
\begin{equation}
    \tilde{A} = 0, \qquad \tilde{\phi} = 0.
\end{equation}
This is not a surprising result since it is also a direct consequence of the Killing equations \eqref{eq:def_killing_equation_nonmetricity} when we use the form with the Levi-Civita covariant derivative. Indeed, this form of the Killing equations is also true in GR, where both the quantities $\tilde{A}$ and $\tilde{\phi}$ are known to be null in a homogeneous LRS spacetime of type II.

\subsection{Flat spatial hypersurfaces}\label{sec:flat_spat_hyper}

Let us consider models that belong to the class of LRS spacetimes of type II characterized by flat spatial hypersurfaces, i.e., we set
\begin{equation}\label{eq:flatness_condition_on_3_Riemann}
    \, ^{3}R^{a}{}_{bcd} = 0, \qquad \, ^{3}\tilde{R}^{a}{}_{bcd} = 0.
\end{equation}

The different combinations of the Gauss relation \eqref{eq:gauss_relation} with $e^{a}$, $N^{ab}$ and $\epsilon^{ab}$, and the flatness conditions \eqref{eq:flatness_condition_on_3_Riemann} and \eqref{eq:lagrange_field_equation}, lead to the following system of equations:
\begin{equation}
\begin{split}
    &\left( 3\Sigma - 2\theta \right)\left( 3\Sigma + 3Q_{7} - 2\theta \right) = 0;\\
    &\left( 3\Sigma - 2\theta \right)\left( 3\Sigma - 3Q_{3} - 2\theta \right) = 0;\\
    &\left( 3\Sigma + \theta \right)\left( 3\Sigma + 3Q_{7} - 2\theta \right) = 0;\\
    &Q_{10}\left( 3\Sigma - 2\theta \right) = 0;\\
    &Q_{10}\left( 3\Sigma + \theta \right) = 0;\\
\end{split}
\end{equation}
whose non-trivial solutions are given by:
\begin{itemize}
    \item solution 1
    \begin{equation}\label{eq:hom_flat_condition_1}
    \left\{ Q_{3} = \Sigma + \frac{1}{3}\theta , \: Q_{7} = -\Sigma + \frac{2}{3}\theta, \: Q_{10} = 0 \right\};
    \end{equation}
    \item solution 2
    \begin{equation}\label{eq:hom_flat_condition_2}
    \left\{ \theta = \frac{3}{2}\Sigma , \: Q_{7} = 0, \: Q_{10} = 0 \right\}.
    \end{equation}
\end{itemize}
We are going to analyze the constraints and equations that can be derived from these two sets of solutions.

\subsubsection{Solution 1}\label{sec:hom_sol_1_system_equat}
From Eq. \eqref{eq:hom_flat_condition_1}, with the help of the Riemann tensor contractions \eqref{eq:riemann_contraction_1}-\eqref{eq:riemann_contraction_13}, the magnetic parts of the Weyl tensor \eqref{eq:magnetic_part_weyl_1}-\eqref{eq:magnetic_part_weyl_5} and the Eq. \eqref{eq:gauss_relation_levi_civita_and_total}, we derive that $\tilde{\theta}$ and $\tilde{\Sigma}$ satisfy the relations
\begin{equation}
    Q_{2} = \frac{2}{3}\left( \tilde{\theta} + 3\tilde{\Sigma} \right), \qquad Q_{6} = \frac{2}{3}\left( 2\tilde{\theta} - 3\tilde{\Sigma} \right),
\end{equation}
and the following sets of constraints for the scalar quantities of the nonmetricity tensor:
\begin{itemize}
    \item constraint 1,
    \begin{equation}\label{eq:sol_1_hom_cons_1}
    \begin{gathered}
    Q_{1} = Q_{4} = Q_{8} = Q_{9} = Q_{10} = Q_{11} = 0,\\
    Q_{5}Q_{3} = 0 \qquad Q_{5}Q_{7} = 0;
    \end{gathered}
    \end{equation}
    \item constraint 2,
    \begin{equation}\label{eq:sol_1_hom_cons_2}
    \begin{gathered}
    Q_{1} = Q_{8} = Q_{9} = Q_{10} = Q_{11} = 0,\\
    Q_{4}=2Q_{5} \qquad Q_{5}Q_{7} = 0.
    \end{gathered}
    \end{equation}
\end{itemize}
Notice that these constraints coincide in the case $Q_5=0$.

Lastly, the Eq. \eqref{eq:levicivita_ricci_uu} and the contraction of the Eq. \eqref{eq:levicivita_contracted_gauss_relation} with the metric tensor lead to the Friedmann equation,
\begin{equation}
    \frac{3}{2} \tilde{\Sigma}^{2} - \frac{2}{3}\tilde{\theta}^{2} +  \frac{2}{f'}\rho + \frac{f}{f'} - \mathcal{Q} + \frac{f''}{f'}(Q_{3} + Q_{7})\dot{Q} = 0.
\end{equation}
By collecting all the previous results, we obtain the following systems of equations:
\begin{itemize}
    \item for the constraint 1 \eqref{eq:sol_1_hom_cons_1} we have
    \begin{equation}\label{eq:hom_final_system1_1_1}
    \begin{split}
    &\dot{{\tilde{\theta}}} + \frac{1}{3}\tilde{\theta}^{2} + \frac{3}{2} \tilde{\Sigma}^{2} + \frac{1}{2f'}\left( \rho + 3 p \right) +\\
    &- \frac{1}{2}\left(\frac{f}{f'} - \mathcal{Q} \right) + \frac{f''}{f'} \tilde{\theta} \dot{\mathcal{Q}} = 0,
    \end{split}
    \end{equation}
    \begin{equation}\label{eq:hom_final_system1_1_2}
    \dot{\tilde{\Sigma}} + \tilde{\theta}\tilde{\Sigma} - \frac{\pi}{f'} - \frac{1}{3}\frac{f''}{f'}\left( 2Q_{3} - Q_{7} - 3\tilde{\Sigma} \right)\dot{\mathcal{Q}} = 0,
    \end{equation}
    \begin{equation}\label{eq:hom_final_system1_1_3}
    q = 0,
    \end{equation}
    \begin{equation}\label{eq:hom_final_system1_1_4}
    \frac{3}{2} \tilde{\Sigma}^{2} - \frac{2}{3}\tilde{\theta}^{2} +  \frac{2}{f'}\rho + \frac{f}{f'} - \mathcal{Q} + \frac{f''}{f'}(Q_{3} + Q_{7})\dot{\mathcal{Q}} = 0,
    \end{equation}
    \begin{equation}\label{eq:hom_final_system1_1_5}
    \tilde{\mathcal{E}} + \frac{1}{2}\tilde{\Sigma}^{2} - \frac{1}{3}\tilde{\theta}\tilde{\Sigma} + \frac{1}{2}\frac{\pi}{f'} + \frac{1}{3}\frac{f''}{f'} \left( Q_{3} - \frac{1}{2} Q_{7} - \frac{3}{2}\tilde{\Sigma} \right)\dot{\mathcal{Q}} = 0
    \end{equation}
    \begin{equation}\label{eq:hom_final_system1_1_6}
        Q_{5}Q_{3} = 0, \qquad Q_{5}Q_{7} = 0,
    \end{equation}
    \begin{equation}\label{eq:hom_final_system1_1_7}
        \dot{Q}_{3} + Q_{3} \left( 2 \tilde{\Sigma} + \frac{2}{3}\tilde{\theta} + \frac{1}{2}Q_{0} \right) = 0,
    \end{equation}
    \begin{equation}\label{eq:hom_final_system1_1_8}
        \dot{Q}_{7} - Q_{7} \left( \tilde{\Sigma} - \frac{2}{3}\tilde{\theta} - \frac{1}{2}Q_{0} \right) = 0,
    \end{equation}
    \begin{equation}\label{eq:hom_final_system1_1_9}
    \begin{split}
    0 =& \tilde{\mathcal{E}} + \frac{1}{2}\dot{\tilde{\Sigma}} + \frac{1}{12}\dot{Q}_{3} - \frac{1}{24}\dot{Q}_{7} + \frac{1}{2}\tilde{\Sigma}^{2} + \frac{1}{6}\tilde{\theta}\tilde{\Sigma} +\\
    &+ \frac{1}{6}\tilde{\Sigma} \left( Q_{3} + \frac{1}{4}Q_{7} \right) + \frac{1}{18} \tilde{\theta} \left( Q_{3} - \frac{1}{2}Q_{7} \right) +\\
    &+ \frac{1}{24}Q_{0}\left( Q_{3} - \frac{1}{2}Q_{7} \right),
    \end{split}    
    \end{equation}
    \begin{equation}\label{eq:hom_final_system1_1_10}
    \begin{split}
        \mathcal{Q} =& -\frac{3}{2}\tilde{\Sigma}^{2} + \frac{2}{3}\tilde{\theta}^{2} + \tilde{\Sigma} \left( 2Q_{3} - Q_{7} \right) +\\
        &- \frac{1}{3}\tilde{\theta}\left( Q_{3} + Q_{7} \right) + \frac{1}{2}Q_{0}\left( Q_{3} + Q_{7}\right);
    \end{split}
    \end{equation}
    
    \item for the constraint 2 \eqref{eq:sol_1_hom_cons_2} the differences with the previous case are in Eqs. \eqref{eq:hom_final_system1_1_3}, \eqref{eq:hom_final_system1_1_6}, \eqref{eq:hom_final_system1_1_7} and \eqref{eq:hom_final_system1_1_9}, with also the addition of a further equation for $Q_{5}$;
    \begin{equation}\label{eq:hom_final_system1_2_3}
    q - \frac{1}{2}\frac{f''}{f'} Q_{5} \dot{\mathcal{Q}}  = 0,
    \end{equation}
    \begin{equation}\label{eq:hom_final_system1_2_6}
        Q_{5}Q_{7} = 0,
    \end{equation}
    \begin{equation}\label{eq:hom_final_system1_2_7}
        \dot{Q}_{3} + Q_{3} \left( 2 \tilde{\Sigma} + \frac{2}{3}\tilde{\theta} + \frac{1}{2}Q_{0} \right) - Q_{5}{}^{2} = 0,
    \end{equation}
    \begin{equation}\label{eq:hom_final_system1_2_9}
    \begin{split}
    0 =& \tilde{\mathcal{E}} + \frac{1}{2}\dot{\tilde{\Sigma}} + \frac{1}{12}\dot{Q}_{3} - \frac{1}{24}\dot{Q}_{7} + \frac{1}{2}\tilde{\Sigma}^{2} -\frac{1}{12}Q_{5}{}^{2}  +\frac{1}{6}\tilde{\theta}\tilde{\Sigma} +\\
    &+ \frac{1}{6}\tilde{\Sigma} \left( Q_{3} + \frac{1}{4}Q_{7} \right) + \frac{1}{18} \tilde{\theta} \left( Q_{3} - \frac{1}{2}Q_{7} \right) +\\
    &+ \frac{1}{24}Q_{0}\left( Q_{3} - \frac{1}{2}Q_{7} \right),
    \end{split}    
    \end{equation}
    \begin{equation}\label{eq:hom_final_system1_2_10}
    \begin{split}
        \mathcal{Q} =& -\frac{3}{2}\tilde{\Sigma}^{2} + \frac{2}{3}\tilde{\theta}^{2} + \tilde{\Sigma} \left( 2Q_{3} - Q_{7} \right) +\\
        &- \frac{1}{3}\tilde{\theta}\left( Q_{3} + Q_{7} \right) + \frac{1}{2}Q_{0}\left( Q_{3} + Q_{7} \right) - Q_{5}{}^{2},
    \end{split}
    \end{equation}
    \begin{equation}\label{eq:hom_final_system1_2_11}
    \dot{Q}_{5} - Q_{5} \left( \tilde{\Sigma} + \frac{1}{3}\tilde{\theta} \right) = 0.
    \end{equation}
\end{itemize}

Equation \eqref{eq:electric_part_weyl_2} is not included in the system of Eqs. \eqref{eq:hom_final_system1_1_1}-\eqref{eq:hom_final_system1_1_10} and \eqref{eq:hom_final_system1_2_3}-\eqref{eq:hom_final_system1_2_11} since it can be proved that the combination of the Eqs. \eqref{eq:electric_part_weyl_1} and \eqref{eq:electric_part_weyl_2} is equal to combining the Eqs. \eqref{eq:hom_final_system1_1_7} and \eqref{eq:hom_final_system1_1_8}.

\subsubsection{Solution 2}\label{sec:hom_sol_2_system_equat}
Following the same procedure described in Sec. \ref{sec:hom_sol_1_system_equat}, we obtain two new sets of constraints for Eq. \eqref{eq:hom_flat_condition_2}:
\begin{itemize}
    \item constraint 1,
    \begin{equation}\label{eq:sol_2_hom_cons_1}
    \begin{gathered}
        Q_{1}=Q_{4}=Q_{7}=Q_{8}=Q_{9}=Q_{10}=Q_{11}=0,\\
        \\
        Q_{2} = \frac{1}{3}\left( 2\tilde{\theta} + 6\tilde{\Sigma} - 9\Sigma + 6Q_{3} \right), \quad Q_{6} = \frac{2}{3}\left( 2\tilde{\theta} - 3\tilde{\Sigma} \right);
    \end{gathered}
    \end{equation}
    \item constraint 2,
    \begin{equation}\label{eq:sol_2_hom_cons_2}
    \begin{gathered}
        Q_{1}=Q_{4}=Q_{7}=Q_{8}=Q_{9}=Q_{10}=Q_{11}=0, \\ 
        \\
        \Sigma = 0,\\
        \\
        Q_{2} = \frac{1}{3}\left( 2\tilde{\theta} + 6\tilde{\Sigma} + 6Q_{3} \right), \quad Q_{6} = \frac{2}{3}\left( 2\tilde{\theta} - 3\tilde{\Sigma} \right); 
    \end{gathered}
    \end{equation}
\end{itemize}
The main difference between the two constraints above is the condition on the shear $\Sigma$ in \eqref{eq:sol_2_hom_cons_2}. Such a condition implies spherical symmetry in the manifold endowed with the full connection, which is present only in \eqref{eq:sol_2_hom_cons_2}. Otherwise, these conditions present minimal differences from the ones in Sec. \ref{sec:hom_sol_1_system_equat}, namely in the roles of $Q_{3}$ and $Q_{7}$.

The two sets of equations to be satisfied are the following:
\begin{itemize}
    \item constraint 1 \eqref{eq:sol_2_hom_cons_1}
    \begin{equation}\label{eq:hom_final_system2_1_1}
    \begin{split}
    &\dot{{\tilde{\theta}}} + \frac{1}{3}\tilde{\theta}^{2} + \frac{3}{2} \tilde{\Sigma}^{2} + \frac{1}{2f'}\left( \rho + 3 p \right) +\\
    &- \frac{1}{2}\left(\frac{f}{f'} - \mathcal{Q} \right) + \frac{f''}{f'} \left( \tilde{\theta} + Q_{3} - \frac{3}{2}\Sigma \right) \dot{\mathcal{Q}} = 0,
    \end{split}
    \end{equation}
    \begin{equation}\label{eq:hom_final_system2_1_2}
    \dot{\tilde{\Sigma}} + \tilde{\theta}\tilde{\Sigma} - \frac{\pi}{f'} - \frac{f''}{f'}\left( \Sigma - \tilde{\Sigma} \right)\dot{\mathcal{Q}} = 0,
    \end{equation}
    \begin{equation}\label{eq:hom_final_system2_1_3}
    q = 0,
    \end{equation}
    \begin{equation}\label{eq:hom_final_system2_1_4}
    \frac{3}{2} \tilde{\Sigma}^{2} - \frac{2}{3}\tilde{\theta}^{2} +  \frac{2}{f'}\rho + \frac{f}{f'} - \mathcal{Q} + \frac{f''}{f'}Q_{3}\dot{\mathcal{Q}} = 0,
    \end{equation}
    \begin{equation}\label{eq:hom_final_system2_1_5}
    \tilde{\mathcal{E}} + \frac{1}{2}\tilde{\Sigma}^{2} - \frac{1}{3}\tilde{\theta}\tilde{\Sigma} + \frac{1}{2}\frac{\pi}{f'} + \frac{1}{2}\frac{f''}{f'} \left( \Sigma - \tilde{\Sigma} \right)\dot{\mathcal{Q}} = 0
    \end{equation}
    \begin{equation}\label{eq:hom_final_system2_1_6}
        Q_{5}\Sigma = 0,
    \end{equation}
    \begin{equation}\label{eq:hom_final_system2_1_7}
        \dot{\Sigma} - \frac{3}{2}\Sigma^{2} + \frac{2}{3}\Sigma \left( 3 \tilde{\Sigma} + \tilde{\theta} + \frac{3}{2}Q_{3} + \frac{3}{4}Q_{0} \right) = 0,
    \end{equation}
    \begin{equation}\label{eq:hom_final_system2_1_8}
    \begin{split}
        &\dot{Q}_{3} - Q_{3}{}^{2} - \frac{9}{2}\Sigma^{2} - \frac{1}{2}Q_{0}Q_{3} +\\
        &+ \Sigma\left( 3\tilde{\Sigma} + \tilde{\theta} + \frac{9}{2}Q_{3} + \frac{3}{2}Q_{0} \right) = 0,
    \end{split}
    \end{equation}
    \begin{equation}\label{eq:hom_final_system2_1_9}
    \begin{split}
    0 =& \tilde{\mathcal{E}} -\frac{1}{2}\dot{\Sigma} + \frac{1}{2}\dot{\tilde{\Sigma}} + \frac{5}{12}\dot{Q}_{3} - \frac{9}{8}\Sigma^{2} + \frac{1}{2}\tilde{\Sigma}^{2} +\\
    &- \frac{5}{12}Q_{3}{}^{2} + \frac{1}{6}\tilde{\theta}\tilde{\Sigma} - \frac{5}{24}Q_{0}Q_{3} +\\
    &+ \Sigma \left( \frac{1}{4}\tilde{\Sigma} + \frac{1}{12}\tilde{\theta} + \frac{11}{8}Q_{3} + \frac{3}{8}Q_{0} \right),
    \end{split}    
    \end{equation}
    \begin{equation}\label{eq:hom_final_system2_1_10}
    \begin{split}
        \mathcal{Q} =& -\frac{3}{2}\tilde{\Sigma}^{2} + \frac{2}{3}\tilde{\theta}^{2} + Q_{3}{}^{2} + Q_{3}\left( \tilde{\theta} + \frac{1}{2}Q_{0} \right) +\\
        &+ \Sigma \left( 3\tilde{\Sigma} - 2\tilde{\theta} - \frac{3}{2}Q_{3}\right).
    \end{split}
    \end{equation}
    
    \item as said above, the constraint 2 \eqref{eq:sol_2_hom_cons_2} differs from \eqref{eq:sol_2_hom_cons_1} by the condition $\Sigma=0$, so we have
    \begin{equation}\label{eq:hom_final_system2_2_1}
    \begin{split}
    &\dot{{\tilde{\theta}}} + \frac{1}{3}\tilde{\theta}^{2} + \frac{3}{2} \tilde{\Sigma}^{2} + \frac{1}{2f'}\left( \rho + 3 p \right) +\\
    &- \frac{1}{2}\left(\frac{f}{f'} - \mathcal{Q} \right) + \frac{f''}{f'} \left( \tilde{\theta} + Q_{3} \right) \dot{\mathcal{Q}} = 0,
    \end{split}
    \end{equation}
    \begin{equation}\label{eq:hom_final_system2_2_2}
    \dot{\tilde{\Sigma}} + \tilde{\theta}\tilde{\Sigma} - \frac{\pi}{f'} + \frac{f''}{f'} \tilde{\Sigma}\dot{\mathcal{Q}} = 0,
    \end{equation}
    \begin{equation}\label{eq:hom_final_system2_2_3}
    q = 0,
    \end{equation}
    \begin{equation}\label{eq:hom_final_system2_2_4}
    \frac{3}{2} \tilde{\Sigma}^{2} - \frac{2}{3}\tilde{\theta}^{2} +  \frac{2}{f'}\rho + \frac{f}{f'} - \mathcal{Q} + \frac{f''}{f'}Q_{3}\dot{\mathcal{Q}} = 0,
    \end{equation}
    \begin{equation}\label{eq:hom_final_system2_2_5}
    \tilde{\mathcal{E}} + \frac{1}{2}\tilde{\Sigma}^{2} - \frac{1}{3}\tilde{\theta}\tilde{\Sigma} + \frac{1}{2}\frac{\pi}{f'} - \frac{1}{2}\frac{f''}{f'} \tilde{\Sigma} \dot{\mathcal{Q}} = 0,
    \end{equation}
    \begin{equation}\label{eq:hom_final_system2_2_8}
    \begin{split}
        &\dot{Q}_{3} - Q_{3}{}^{2} - \frac{1}{2}Q_{0}Q_{3} = 0,
    \end{split}
    \end{equation}
    \begin{equation}\label{eq:hom_final_system2_2_9}
    \begin{split}
    0 =& \tilde{\mathcal{E}} + \frac{1}{2}\dot{\tilde{\Sigma}} + \frac{5}{12}\dot{Q}_{3} + \frac{1}{2}\tilde{\Sigma}^{2} +\\
    &- \frac{5}{12}Q_{3}{}^{2} + \frac{1}{6}\tilde{\theta}\tilde{\Sigma} - \frac{5}{24}Q_{0}Q_{3},
    \end{split}    
    \end{equation}
    \begin{equation}\label{eq:hom_final_system2_2_10}
    \begin{split}
        \mathcal{Q} =& -\frac{3}{2}\tilde{\Sigma}^{2} + \frac{2}{3}\tilde{\theta}^{2} + Q_{3}{}^{2} + Q_{3}\left( \tilde{\theta} + \frac{1}{2}Q_{0} \right).
    \end{split}
    \end{equation}
\end{itemize}
Once again, the Eq. \eqref{eq:electric_part_weyl_2} is not included in the system of equations for the same reasons given in Sec. \ref{sec:hom_sol_1_system_equat}

\subsection{FLRW type metrics}
As an application of what has been done in the above Sec. \ref{sec:flat_spat_hyper}, we consider the spatially flat FLRW type metric among the several classes of metrics with flat spatial hypersurfaces. They are characterized by the requirement
\begin{equation}\label{eq:FLRW_condition}
    \tilde{\Sigma} = 0,
\end{equation}
and the energy-momentum tensor of a perfect fluid,
\begin{equation}
    \Psi_{ab} = \rho u_{a}u_{b} + p h_{ab}.
\end{equation}
Because of the Eq. \eqref{eq:energy-momentum_conservation}, we have that the usual energy-momentum conservation law is true,
\begin{equation}
    \dot{\rho} + \tilde{\theta}\left( \rho + p \right) = 0.
\end{equation}
In coordinates, spatially flat FLRW type metrics are represented as follows,
\begin{equation}\label{eq:FLRW_metric}
    \text{d}s^{2} = - \mathbb{N}(t)^{2}\text{d}t^{2} + a(t)^{2} \left( \text{d}x^{2} + \text{d}y^{2} + \text{d}z^{2} \right),
\end{equation}
where $\mathbb{N}(t)$ is the time lapse function and $a(t)$ is the scale factor. We can also introduce the relation\footnote{Notice that the coordinate representation is not an essential requirement to introduce the relation \eqref{eq:rel_scal_factor_theta} between the scale factor $a$ and the expansion rate $\tilde{\theta}$.}
\begin{equation}\label{eq:rel_scal_factor_theta}
    \frac{\dot{a}}{a} = \frac{1}{3}\tilde{\theta},
\end{equation}
and the $4$-velocity $u^{a}$ and the spatial $4$-vector $e^{a}$ expressed as
\begin{equation}
    u^{a} \equiv \bigg\{ \frac{1}{\mathbb{N}}, \: 0, \: 0, \: 0 \bigg\}, \quad \text{and} \quad e^{a} \equiv \bigg\{ 0, \: \frac{1}{a}, \: 0, \: 0 \bigg\}.
\end{equation}
Therefore, the time derivative of a generic scalar $\psi(t)$ is equal to
\begin{equation}
    \dot{\psi} = u^{a}\nabla_{a}\psi = u^{a}\partial_{a}\psi = \frac{1}{\mathbb{N}}\partial_{t}\psi.
\end{equation}
 Since we have $\tilde{A}=0$, we obtain that the lapse function is constant, and we can set it to $1$ without loss of generality:
\begin{equation}
    \mathbb{N}(t) = const. = 1.
\end{equation}

Applying the condition \eqref{eq:FLRW_condition} to the systems of equations derived in Secs. \ref{sec:hom_sol_1_system_equat}, we find three different solutions:
\begin{itemize}
    \item solution 1 of the system \eqref{eq:hom_final_system1_1_1}-\eqref{eq:hom_final_system1_1_10},
    \begin{equation}\label{eq:FLRW_Sol_1}
    \begin{gathered}
        Q_{3}=Q_{5}=Q_{7}=0, \qquad Q_{6} = 2 Q_{2} = \frac{4}{3}\tilde{\theta},\\
        \\
        \Sigma=\theta=0,\\
        \\
        \Gamma_{00}{}^{0} = \frac{1}{2}Q_{0};
    \end{gathered} 
    \end{equation}
    \item solution 2 of the system \eqref{eq:hom_final_system1_1_1}-\eqref{eq:hom_final_system1_1_10},
    \begin{equation}\label{eq:FLRW_Sol_2}
    \begin{gathered}
        Q_{3}=Q_{7}=0, \quad Q_{5}\neq0 \quad Q_{6} = 2 Q_{2} = \frac{4}{3}\tilde{\theta},\\
        \\
        \Sigma=\theta=0,\\
        \\
        \Gamma_{00}{}^{0} = \frac{1}{2}Q_{0} \qquad \Gamma_{00}{}^{1} = - \frac{Q_{5}}{a};
    \end{gathered} 
    \end{equation}
    \item solution 3 of both systems \eqref{eq:hom_final_system1_1_1}-\eqref{eq:hom_final_system1_1_10} and \eqref{eq:hom_final_system1_2_3}-\eqref{eq:hom_final_system1_2_11},
    \begin{equation}\label{eq:FLRW_Sol_3}
    \begin{gathered}
        Q_{5} = 0, \quad Q_{3} = \frac{Q_{7}}{2}, \quad Q_{6} = 2 Q_{2} = \frac{4}{3}\tilde{\theta},\\
        \\
        \Sigma = 0, \qquad \theta \neq 0,\\
        \\
        \Gamma_{00}{}^{0} = \frac{1}{2}Q_{0} \qquad \Gamma_{11}{}^{0} = \Gamma_{22}{}^{0} = \Gamma_{33}{}^{0} = \frac{1}{2}a^{2}Q_{7}.
    \end{gathered} 
    \end{equation}
\end{itemize}
On the other hand, for the systems of equations in Sec. \ref{sec:hom_sol_2_system_equat}, we have:
\begin{itemize}
    \item solution 4 of both systems \eqref{eq:hom_final_system2_1_1}-\eqref{eq:hom_final_system2_1_10} and \eqref{eq:hom_final_system2_2_1}-\eqref{eq:hom_final_system2_2_10},
    \begin{equation}\label{eq:FLRW_Sol_4}
    \begin{gathered}
        Q_{7} = 0, \quad Q_{3} \neq 0, \quad Q_{5} \neq 0 \quad Q_{2} = \frac{1}{2}Q_{6} + 2 Q_{3},\\
        \\
        \Sigma = \theta = 0,\\
        \\
        \Gamma_{00}{}^{0} = \frac{1}{2}Q_{0}, \quad \Gamma_{00}{}^{1} = - \frac{Q_{5}}{a}, \quad \Gamma_{01}{}^{1} = - Q_{3}.
    \end{gathered} 
    \end{equation}
\end{itemize}
The components of the total connection that are not represented are zero.

All the above results are obtained by excluding the two conditions $f'' = 0$ and  $\mathcal{Q} = const.$ since, as is known, they always lead back to GR with a cosmological constant.

As an example, we take the result \eqref{eq:FLRW_Sol_1} and set $Q_{0}=0$ for the sake of simplicity\footnote{With this choice, we obtain a total connection that is null, that is, we have the coincident gauge.}. Moreover, we also require that $p$ and $\rho$ satisfy the barotropic linear equation of state,
\begin{equation}
p = w\rho, \qquad w=const.
\end{equation}
The remaining equations to be solved are
\begin{equation}
    \mathcal{E} = 0,
\end{equation}
\begin{equation}
    \mathcal{Q} = \frac{2}{3}\tilde{\theta}^{2},
\end{equation}
\begin{equation}
\begin{split}
    &\dot{{\tilde{\theta}}} + \frac{1}{3}\tilde{\theta}^{2} + \frac{1}{2f'}\left( 1 + 3 w \right)\rho +\\
    &- \frac{1}{2}\left(\frac{f}{f'} - \mathcal{Q} \right) + \frac{f''}{f'} \tilde{\theta} \dot{\mathcal{Q}} = 0,
\end{split}
\end{equation}
\begin{equation}
    - \frac{2}{3}\tilde{\theta}^{2} +  \frac{2}{f'}\rho + \frac{f}{f'} - \mathcal{Q} = 0,
\end{equation}
\begin{equation}
    \dot{\rho} + \tilde{\theta}\left( 1 + w \right)\rho = 0.
\end{equation}
To find a solution to the above equations, we consider a power-law function for $f(\mathcal{Q})$,
\begin{equation}
    f(\mathcal{Q}) = \alpha \mathcal{Q}^{n},
\end{equation}
with $\alpha$ a generic dimensional constant. Hence, the solution turns out to be
\begin{equation}
    \tilde{\theta} = \frac{2n}{\left( 1 + w \right)t - 2n \mathbb{C}_{1}},
\end{equation}
\begin{equation}
    \rho = \alpha \left( 2n - 1 \right) \frac{2^{-1+3n}}{3^{n}} \left[ \frac{n}{\left( 1 + w \right)t - 2n \mathbb{C}_{1}} \right]^{2n},
\end{equation}
\begin{equation}
    a = \mathbb{C}_{2}\left[ \left( 1 + w \right)t - 2n \mathbb{C}_{1} \right]^{\frac{2n}{3(1+w)}},
\end{equation}
where $\mathbb{C}_{1}$ and $\mathbb{C}_{2}$ are constants of integration.

The relevance of the results given here lies in the fact that the solutions \eqref{eq:FLRW_Sol_1}-\eqref{eq:FLRW_Sol_4} have a different form of the connections from those already shown in the literature, where both the metric and the connection are assumed invariant under rotations and spatial translations, such as \cite{DAmbrosio:2021pnd,Dimakis:2022rkd}.

\section{Static spacetime}\label{sec:static_spacetime}
The spacetime is said ``stationary'' if it admits a timelike Killing vector field $\mathbf{Y}$,
\begin{equation}\label{eq:stat_lie_derivative_metric_1_1_2}
    \mathcal{L}_{\mathbf{Y}} g_{ab} = 0.
\end{equation}
Once again, we also require the condition that the Lie derivative of the nonmetricity with respect to $\mathbf{Y}$ is zero,
\begin{equation}\label{eq:stat_lie_derivative_nonmetricity_1_1_2}
    \mathcal{L}_{\mathbf{Y}} Q_{abc} = 0.
\end{equation}

Moreover, if the congruence associated with $\mathbf{Y}$ is hypersurface orthogonal, the spacetime is said ``static''. For this reason, we consider the Killing vector
\begin{equation}
    \mathbf{Y} = \chi(\mathbf{x})u,
\end{equation}
where $\chi(\mathbf{x})$ is a generic smooth function of the coordinates. In this way, given the condition \eqref{eq:hyper_ortho_u_e}, we have that $\mathbf{Y}$ is hypersurface orthogonal since it is proportional to the $4$-velocity, and so our spacetime is static. 

Considering the same contractions of the Eq. \eqref{eq:def_killing_equation_nonmetricity} we used in Sec. \ref{sec:hom_spacetime}, we obtain conditions on $\chi$ and the kinematic quantities. The double contraction with $u^{a}$ gives us,
\begin{equation}
    \dot{\chi} = 0,
\end{equation}
that is, the function $\chi$ is independent of time.
Instead, the contraction with $u^{a}$ and $e^{b}$ leads to
\begin{equation}
    \hat{\chi} = \tilde{\mathcal{A}} \chi.
\end{equation}
Moreover, contracting twice with $e^{a}$ and with $N^{ab}$ yield
\begin{equation}\label{eq:lie_static_ee}
    \Sigma + \frac{1}{3}\theta + \frac{1}{2}Q_{2} - Q_{3} = 0,
\end{equation}
\begin{equation}\label{eq:lie_theta_sigma}
\begin{gathered}
    \theta = -\frac{1}{2}Q_{2} + Q_{3} - \frac{1}{2}Q_{6} + Q_{7}, \\
    \Sigma = -\frac{1}{3}Q_{2} + \frac{2}{3}Q_{3} + \frac{1}{6}Q_{6} - \frac{1}{3} Q_{7}.
\end{gathered}
\end{equation}

From Eq. \eqref{eq:stat_lie_derivative_nonmetricity_1_1_2}, we have instead that the time derivative of the scalar components of the nonmetricity tensor vanishes:
\begin{equation}
    \dot{Q}_{i} = 0, \qquad i=0,...,11.
\end{equation}
Therefore, we can assume that the covariant derivatives of the scalars \eqref{LRS_scalars} are zero except for those along the radial direction $e^{a}$.

As happens for the Eq. \eqref{eq:hom_cond_Au} in the homogeneous case, we can highlight that the relations \eqref{eq:lie_static_ee} can also be derived from the commutation relation \eqref{eq:commutation_relation} by imposing, this time, that the time derivative is null.

Taking in mind the Eq. \eqref{eq:theta_sigma_phi_A_levicivita_nonmetricity}, because of the Eq. \eqref{eq:lie_theta_sigma} we have the identities
\begin{equation}\label{eq:stat_levi_civita_theta_sigma}
    \tilde{\theta}=0, \qquad \text{and} \qquad \tilde{\Sigma}=0.
\end{equation}
Again, this is an expected result since $\tilde{\theta}$ and $\tilde{\Sigma}$ are null in GR when we consider a stationary LRS spacetime of type II.

Equations \eqref{eq:lie_theta_sigma} and \eqref{eq:stat_levi_civita_theta_sigma} mean that in the presence of a non-metric connection, expansion and shear can manifest exclusively due to the nonmetricity tensor. Therefore, a static metric is spherically symmetric, as in GR.

\subsection{The covariant equations for static spacetimes}
Some preliminary considerations are in order before writing the final system of equations to be solved. 

First, we are looking for a vacuum solution, so $\rho=p=q=\pi=0$. Moreover, we require that the curves of the timelike congruence are autoparallel\footnote{Such a choice is made to simplify the mathematics.}, i.e., $u^{c}\nabla_{c}u^{a} = 0$, which implies
\begin{equation}\label{eq:cond_autoparallelism}
    Q_{0} = 0, \quad \mathcal{A}^{(e)} = 0, \quad \text{and} \quad \mathcal{A}^{(u)} = Q_{5}.
\end{equation}
Equations \eqref{eq:cond_autoparallelism} and  \eqref{eq:levicivita_ricci_ue} give rise to the identity
\begin{equation}\label{eq:stat_cond_Q2_Q6}
    Q_{2}=Q_{6},
\end{equation}
if we exclude the two conditions $f'' = 0$ and  $\mathcal{Q} = const.$

Moreover, the Eqs. \eqref{eq:cond_autoparallelism} and \eqref{eq:stat_cond_Q2_Q6}, together with the Riemann tensor contractions \eqref{eq:riemann_contraction_1}-\eqref{eq:riemann_contraction_13} and the magnetic parts of Weyl tensor \eqref{eq:magnetic_part_weyl_1}-\eqref{eq:magnetic_part_weyl_5}, lead to the following constraints,
\begin{equation}
    Q_{6} = Q_{10} = Q_{11} = 0,  \quad Q_{3}Q_{5} = 0, \quad Q_{5}Q_{7} = 0,
\end{equation}
\begin{equation}
    \tilde{A} = - \frac{1}{2}Q_{4}.
\end{equation}
Hence, we can write the final system of field equations:
\begin{equation}\label{eq:fse_1}
    \hat{\tilde{\mathcal{A}}} + \tilde{\mathcal{A}}^{2} + \tilde{\mathcal{A}}\tilde{\phi} + \frac{1}{2}\frac{f}{f'} - \frac{1}{2}\mathcal{Q} - \frac{1}{2}\frac{f''}{f'}\left( Q_{4} - 2Q_{5} \right)\hat{\mathcal{Q}} = 0,
\end{equation}
\begin{equation}\label{eq:fse_2}
\begin{split}
    \tilde{\mathcal{E}} + \tilde{\mathcal{A}}\tilde{\phi} &+ \frac{1}{3}\frac{f}{f'} - \frac{1}{3}\mathcal{Q} - \frac{1}{6}\frac{f''}{f'}Q_{4}\hat{\mathcal{Q}} + \frac{2}{3}\frac{f''}{f'}Q_{5}\hat{\mathcal{Q}} +\\
    &\qquad \qquad - \frac{1}{12}\frac{f''}{f'}Q_{8}\hat{\mathcal{Q}} - \frac{1}{6}\frac{f''}{f'}Q_{9}\hat{\mathcal{Q}} = 0,
\end{split}
\end{equation}

\begin{equation}\label{eq:fse_3}
    \hat{\tilde{\phi}} + \frac{1}{2}\tilde{\phi}^{2} - \tilde{\mathcal{A}}\tilde{\phi} - \frac{f''}{f'}Q_{5}\hat{\mathcal{Q}} + \frac{1}{2}\frac{f''}{f'}Q_{8} \hat{\mathcal{Q}}  = 0,
\end{equation}

\begin{equation}\label{eq:fse_4}
    \hat{Q}_{5} + Q_{5}{}^{2} + \frac{1}{2}Q_{1}Q_{5} = 0,
\end{equation}

\begin{equation}\label{eq:fse_5}
    \hat{\phi} -  \hat{Q}_{9} + \frac{\phi ^{2}}{2} + \frac{Q_{9}^{2}}{2} - \phi  \left(Q_{9}-\frac{Q_{1}}{2}\right) - \frac{1}{2} Q_{1}Q_{9}  = 0,
\end{equation}

\begin{equation}\label{eq:fse_6}
    Q_{5}\phi = 0,
\end{equation}

\begin{equation}\label{eq:fse_7}
    \hat{\phi} + \frac{1}{2}\phi^{2} - \frac{1}{2}\left( Q_{1} - Q_{8} + Q_{9} \right)\phi = 0,
\end{equation}
\begin{equation}\label{eq:fse_8}
    \hat{Q}_{7} - \phi \left( Q_{3} - \frac{1}{2}Q_{7} \right) + \frac{1}{2}Q_{7}\left( Q_{4} + Q_{8} - Q_{9} \right) = 0,
\end{equation}
\begin{equation}\label{eq:fse_9}
    Q_{3}Q_{5} = 0, \qquad Q_{5}Q_{7} = 0,
\end{equation}
\begin{equation}\label{eq:fse_10}
\begin{split}
    0 =& \tilde{\mathcal{E}} + \frac{1}{6}\hat{Q}_{4} - \frac{5}{12}\hat{Q}_{5} + \frac{1}{12}\hat{Q}_{8} - \frac{5}{24}\hat{Q}_{9} - \frac{3}{32}Q_{8}Q_{9} +\\
    & - \frac{Q_{4}^{2}}{12} + \frac{5Q_{8}^{2}}{96} + Q_{5} \left( \frac{5Q_{4}}{12} - \frac{5Q_{8}}{48} \right) +\\
    &+ \tilde{\phi}\left( \frac{Q_{1}}{8} - \frac{Q_{4}}{12} + \frac{5Q_{5}}{24} - \frac{5Q_{8}}{48} - \frac{Q_{9}}{48} \right) +\\
    &+ Q_{1}\left( \frac{5Q_{5}}{24} - \frac{Q_{8}}{16} + \frac{5Q_{9}}{48} \right),
\end{split}    
\end{equation}
\begin{equation}\label{eq:fse_11}
\begin{split}
    0 =& -\tilde{\mathcal{E}} - \frac{1}{6}\hat{Q}_{4} - \frac{1}{12}\hat{Q}_{5} - \frac{1}{12}\hat{Q}_{8} - \frac{1}{24}\hat{Q}_{9} + \frac{1}{32}Q_{8}Q_{9} +\\
    & + \frac{Q_{4}^{2}}{12} + \frac{Q_{8}^{2}}{96} + Q_{5} \left( \frac{Q_{4}}{12} - \frac{Q_{8}}{48} \right) +\\
    &+ \tilde{\phi}\left( \frac{Q_{1}}{8} + \frac{Q_{4}}{12} + \frac{Q_{5}}{24} - \frac{Q_{8}}{48} - \frac{5Q_{9}}{48} \right) +\\
    &+ Q_{1}\left(  \frac{Q_{5}}{24} - \frac{Q_{8}}{16} + \frac{Q_{9}}{48} \right),
\end{split}
\end{equation}
\begin{equation}\label{eq:fse_12}
\begin{split}
    &\hat{Q}_{7} -\tilde{\phi}\left( Q_{3} - \frac{Q_{7}}{2} \right) - Q_{3} \left( \frac{12Q_{5}}{5} - \frac{Q_{8}}{2} + Q_{9} \right) +\\
    &+ Q_{7}\left( \frac{Q_{4}}{2} + \frac{Q_{5}}{5} + \frac{Q_{8}}{4} \right) = 0,
\end{split}
\end{equation}
\begin{equation}\label{eq:fse_13}
\begin{split}
    \mathcal{Q} =& \frac{1}{2}Q_{1}\left( -Q_{5} + Q_{9} \right) - \frac{1}{2} Q_{4} \left( Q_{5} - Q_{8} + Q_{9} \right) +\\
    &- \frac{1}{8} Q_{8} \left(4Q_{5} + Q_{8}\right).
\end{split}
\end{equation}


\subsection{Spherically symmetric solutions}\label{sec:solutions}
In this section, we specialize the system of covariant equations \eqref{eq:fse_1}-\eqref{eq:fse_13} in a spherically symmetric scenario. For this purpose, we introduce the following expression of a static spherically symmetric metric,
\begin{equation}\label{eq:sph_metric}
    \text{d}s^{2} = - \mathbb{A}(r)\text{d}t^{2} + \mathbb{B}(r)\text{d}r^{2} + r^{2} \left( \text{d}\Theta^{2} + \sin^{2}{\Theta}\text{d}\varphi^{2}\right),
\end{equation}
where $\mathbb{A}(r)$ and $\mathbb{B}(r)$ are generic positive functions of the radial coordinate $r$. In this coordinate system, the $4$-velocity $u^{a}$ and the spatial $4$-vector $e^{a}$ are expressed as
\begin{equation}\label{eq:def_u_e_spherical_coordinates}
    u^{a} \equiv \bigg\{ \frac{1}{\sqrt{\mathbb{A}}}, \: 0, \: 0, \: 0 \bigg\}, \quad \text{and} \quad e^{a} \equiv \bigg\{ 0, \: \frac{1}{\sqrt{\mathbb{B}}}, \: 0, \: 0 \bigg\}.
\end{equation}
Using the Eq. \eqref{eq:def_u_e_spherical_coordinates} and the definitions of $\tilde{\phi}$ and $\tilde{\mathcal{A}}$, we obtain the relations,
\begin{equation}\label{eq:metric_00}
    \frac{\mathbb{A}'(r)}{\mathbb{A}(r)}=\frac{4 \tilde{\mathcal{A}}(r)}{r \tilde{\phi} (r)},
\end{equation}
\begin{equation}\label{eq:metric_11}
    \mathbb{B}(r) = \frac{4}{r^{2}}\frac{1}{\tilde{\phi}(r)^{2}},
\end{equation}
\begin{equation}
    e^{a}\nabla_{a}\psi = e^{a}\tilde{\nabla}_{a}\psi = e^{a}\partial_{a}\psi = \frac{1}{2}\tilde{\phi}\: r\: \partial_{r} \psi,
\end{equation}
where $\psi$ is an arbitrary scalar function.
From Eq. \eqref{eq:fse_6} two conditions arise: 
\begin{equation}\label{eq:conditions}
    \phi = 0 \qquad \text{or} \qquad Q_{5}=0.
\end{equation}
To test the viability of both the solutions \eqref{eq:conditions}, we consider the function $f(\mathcal{Q})=\mathcal{Q}$ and verify that, in this case, the equations yield the Schwarzschild spacetime as a solution. The condition $\phi=0$ leads to an unacceptable result for $f(\mathcal{Q})=\mathcal{Q}$: all kinematic scalars are null. For this reason, we discard this branch of solutions, and we narrow down our investigation to the branch $Q_{5}=0$. In such a circumstance, we easily obtain the following relations for $\tilde{\phi}$ and $\tilde{\mathcal{A}}$ when $f(\mathcal{Q})=\mathcal{Q}$,
\begin{equation}
    \tilde{\phi} = \frac{2}{r}\sqrt{1-\frac{r_{s}}{r}}, \qquad \tilde{\mathcal{A}} = -\frac{1}{2} Q_{4} = \frac{r_{s}}{2 r \sqrt{r^{2}-rr_{s}}},
\end{equation}
which substituted into the Eqs. \eqref{eq:metric_00} and \eqref{eq:metric_11} give us the Schwarzschild metric as solution,
\begin{equation}
    \mathbb{A} = \mathbb{B}^{-1} =  1 - \frac{r_{s}}{r},
\end{equation}
being $r_{s}$ the Schwarzschild radius, as expected. 

Moreover, by choosing $Q_{5} = 0$, we have that the Eqs. \eqref{eq:fse_8} and \eqref{eq:fse_12} are equal, and the sum of Eqs. \eqref{eq:fse_10} and \eqref{eq:fse_11} is given by a combination of the Eqs. \eqref{eq:fse_4}-\eqref{eq:fse_8}. Therefore, we do not consider Eq. \eqref{eq:fse_11} in the resolution of our system of equations.

\subsection{Schwarzschild-de Sitter type solutions}\label{sec:Schwarzschild-de Sitter solutions}
In order for the Eqs. \eqref{eq:metric_00} and \eqref{eq:metric_11}  to admit solutions of Schwarzschild-de Sitter type, we must have that 
\begin{equation}\label{eq:GR_equations}
    \hat{\tilde{\phi}} + \frac{1}{2}\tilde{\phi}^{2} - \tilde{\mathcal{A}}\tilde{\phi}=0,
\end{equation}
i.e., the  GR equations for Schwarzschild-de Sitter must be satisfied. Consequently, the Eq. \eqref{eq:fse_3} and the condition $Q_{5}=0$ imply
\begin{equation}\label{eq:SdS_condition}
    f'' Q_{8} \hat{\mathcal{Q}}  = 0.
\end{equation}
As we already know, the Eq. \eqref{eq:SdS_condition} tells us that to have a Schwarzschild-de Sitter metric as a solution, either the function $f(\mathcal{Q})$ must be linear or $\mathcal{Q}$ must be constant. But there is also a third possibility given by the requirement $Q_{8}=0$.

Now, we can solve the final system of equations with the constraint $Q_{8}=0$. For the sake of simplicity, we also require $Q_{3}=0$. From the Eqs. \eqref{eq:fse_3}, \eqref{eq:fse_5} and \eqref{eq:fse_7} we get the relation
\begin{equation}
    Q_{1} = Q_{4},
\end{equation}
which once substituted into the Eq. \eqref{eq:fse_13} gives us $\mathcal{Q}=0$. The vanishing of the nonmetricity scalar implies
\begin{equation}
\begin{gathered}
    f(\mathcal{Q})\vert_{\mathcal{Q}=0} = f_{0}, \qquad f'(\mathcal{Q})\vert_{\mathcal{Q}=0} = f'_{0}, \\
    f''(\mathcal{Q})\vert_{\mathcal{Q}=0} = f''_{0},
\end{gathered}
\end{equation}
i.e., the function $f$ and its derivatives are constant on-shell. The remaining equations admit the solution
\begin{equation}
\begin{gathered}
    \tilde{\phi} = \frac{2}{r}\sqrt{1 - \frac{r_{s}}{r}-\frac{1}{6} \frac{f_{0}}{f'_{0}} r^{2}}, \\ 
    \tilde{\mathcal{A}} = -\frac{1}{2}Q_{4} = \frac{3 f'_{0} r_{s}-f_{0} r^3}{6 r^{2} \sqrt{{f'_{0}}^{2} \left(1 - \frac{r_{s}}{r}-\frac{1}{6} \frac{f_{0}}{f'_{0}} r^{2}\right)}},
\end{gathered}
\end{equation}
\begin{equation}
    Q_{9} = \frac{6 f'_{0} r_{s}+f_{0} r^{3}}{ 3 r^{2} \sqrt{{f'_{0}}^{2} \left(1 - \frac{r_{s}}{r}-\frac{1}{6} \frac{f_{0}}{f'_{0}} r^{2}\right)} }, \quad 
    \tilde{\mathcal{E}} = -\frac{r_{s}}{r^{3}},
\end{equation}
\begin{equation}\label{eq:sch_Sigma}
\begin{split}
    \Sigma =& -\frac{1}{3}\theta = -\frac{1}{3}Q_{7} =\\
    =& -\frac{1}{3}\frac{1}{r} \mathbb{C}_{4} \sqrt{-12 f_{0}' \left( 1 - \frac{r_{s}}{r} - \frac{1}{6}\frac{f_{0} r^{2}}{f'_{0}} \right)};
\end{split}
\end{equation}
\begin{equation}\label{eq:sch_dS_metric}
    \mathbb{A} = \mathbb{B}^{-1} = 1 - \frac{r_{s}}{r} - \frac{1}{6}\frac{f_{0} r^{2}}{f'_{0}}.
\end{equation}
which represents a Schwarzschild-de Sitter type spacetime. In order that the Eq. \eqref{eq:sch_Sigma} is in agreement with the condition $\mathbb{A}>0$, if $f_{0}' > 0$ the integration constant $\mathbb{C}_{4}$ must assume the values
\begin{equation}
    \mathbb{C}_{4} = 0 \qquad \text{or} \qquad \mathbb{C}_{4} = i.
\end{equation}

For $\mathbb{C}_{4} = 0$, we derive the following components of the total connection:
\begin{equation}\label{eq:total_connection_sol_1}
    \begin{gathered}
        \Gamma_{12}{}^{2} = \Gamma_{13}{}^{3} = \frac{1}{r}, \quad \Gamma_{22}{}^{1} =  -r, \quad \Gamma_{23}{}^{3} = \cot{\Theta}, \\
        \Gamma_{33}{}^{1} = - r \sin^{2}{\Theta}, \quad \Gamma_{33}{}^{2} = -\cos{\Theta}\sin{\Theta}.
    \end{gathered}
\end{equation}
On the other hand, for $\mathbb{C}_{4} = i$, we have two additional components, i.e.,
\begin{equation}\label{eq:total_connection_sol_2}
    \begin{gathered}
        \Gamma_{22}{}^{0} = r \sqrt{3f'_{0}}, \qquad \Gamma_{33}{}^{0} = r \sqrt{3f'_{0}}\sin^{2}{\Theta}.
    \end{gathered}
\end{equation}
We have a similar result for $f_{0}'<0$,
\begin{equation}\label{eq:total_connection_sol_3}
    \begin{gathered}
        \Gamma_{22}{}^{0} = \mathbb{C}_{4} r \sqrt{-3f'_{0}}, \qquad \Gamma_{33}{}^{0} = \mathbb{C}_{4} r \sqrt{-3f'_{0}}\sin^{2}{\Theta},
    \end{gathered}
\end{equation}
with $\mathbb{C}_{4}\in \mathbb{R}$.
Therefore, we have found solutions characterized by the same metric \eqref{eq:sch_dS_metric}, but with different connections, Eqs. \eqref{eq:total_connection_sol_1}, \eqref{eq:total_connection_sol_2} and \eqref{eq:total_connection_sol_3}.

The conditions $Q_{3}=Q_{7}=Q_{8}=0$ and the total connection \eqref{eq:total_connection_sol_1} perfectly match the gauge choice and results obtained in other works (e.g., \cite{Zhao:2021zab, Lin:2021uqa}) in which static spherically symmetric spacetimes are studied, and the full connection is assumed to coincide with the Levi-Civita one of a Minkowski spacetime in spherical coordinates. Moreover, it is easy to show that the solutions \eqref{eq:total_connection_sol_1}, \eqref{eq:total_connection_sol_2}, and \eqref{eq:total_connection_sol_3} represent a subset of the Solution set 2 given in \cite{DAmbrosio:2021zpm} for $c=k=0$.

\subsection{\texorpdfstring{$\mathcal{Q}$}{}-Gravastars} \label{sec:gravastar}
Among the exotic objects proposed as alternatives for black holes there are the so-called gravitational vacuum condensate stars or Gravastars  \cite{Mazur:2001fv,Visser_2004}. A gravastar is essentially a compact object made of a dark energy condensate. Because the nonmetricity terms in equations \eqref{eq:final_field_equation} can be considered as an effective fluid which can have negative pressure, one might ask if $f(\mathcal{Q})$ gravity can admit gravastar solutions without invoking explicitly the presence of a cosmological constant. We dub these solutions ``Q-gravastars'' and give their most straightforward realization in the following.

First, let us show that the $\mathcal{Q}$ dependent terms in  \eqref{eq:final_field_equation} can generate an effective cosmological constant.  Rewriting these equations in the form
\begin{equation}
    \tilde{R}_{ab} -\frac{1}{2}g_{ab}\tilde{R} = \frac{1}{f'}T_{ab} -\frac{1}{2}g_{ab} \left( \frac{f}{f'} - \mathcal{Q} \right) -2 \frac{f''}{f'}P^{c}{}_{ab}\partial_{c}\mathcal{Q}  ,
\end{equation}
we have that, in the case $\mathcal{Q}=\mathcal{Q}_{*}=const.$, the effective energy-momentum tensor,
\begin{equation}
    T^{eff}_{ab} = -\frac{1}{2}g_{ab} \left( \frac{f(\mathcal{Q}_{*})}{f'(\mathcal{Q}_{*})} - \mathcal{Q}_{*} \right)=-g_{ab}\Lambda_{*},
\end{equation}
can be thought of as a fluid characterized by a negative pressure,
\begin{equation}
    \frac{1}{3} T^{eff}_{ab} h^{ab} = p_{eff} = -\rho_{eff} = - T^{eff}_{ab} u^{a}u^{b},
\end{equation}
if 
\begin{equation}
 \Lambda_{*}=\frac{1}{2}\left(\frac{f(\mathcal{Q}_{*})}{f'(\mathcal{Q}_{*})} - \mathcal{Q}_{*}\right)>0,
\end{equation}
which corresponds to a function $f$, which grows slower than linear. It is now clear that, with our assumptions, considering a Schwarzschild radius equal to zero, the solution of the field equations will be just the de Sitter solution,
\begin{equation}\label{eq:dS_metric}
    \mathbb{A} = \mathbb{B}^{-1} = 1 - \frac{1}{3}\Lambda_{*} r^{2}.
\end{equation}

A critical aspect of the Gravastar models is their compatibility with a vacuum (Schwarzschild) exterior. Such compatibility is obtained by the well-known Israel-Darmois junction conditions \cite{israel1966singular} (see also \cite{choquet1982analysis} and \cite{taub1980space}), which we briefly summarize for our specific case in the 1+1+2 language (see \cite{Vignolo2024oea} for additional details on the strategy with nonmetricity, and \cite{Rosa:2023tb} for a more general treatment of 1+1+2 junction conditions).

Let $\mathcal{S}$ be a $3$-dimensional hypersurface, with a spacelike normal $n^{a}$, separating the spacetime in two regions denoted by $\mathcal{M}^{+}$ and $\mathcal{M}^{-}$, in which Schwarzschild and de Sitter solutions hold respectively. Then, in the $1+1+2$ framework, we can choose the vector $e^{a}$ as equal to the normal
vector $n^{a}$, and the induced metric on the boundary will be given by
\begin{equation}
    q_{ab}=N_{ab} - u_{a}u_{b}=g_{ab} - e_{a}e_{b}.
\end{equation} 
Assuming smooth $e^a$ and $N_{ab}$, and indicating with $\tilde{\mathcal{K}}_{ab}$ the Levi-Civita extrinsic curvature of $\mathcal{S}$, the standard Israel-Darmois junction conditions can be formulated as
\begin{align}
    \left[q_{ab}\right] &= 0\label{eq:junction_conditions_q}, \\
    \left[ \tilde{\mathcal{K}}_{ab} \right] &= 0
    \label{eq:junction_conditions_K},
\end{align}
where 
\begin{equation}
\begin{split}
    \tilde{\mathcal{K}}_{ab} =& \left( N_{a}{}^{c} - u_{a}u^{c} \right)\left( N_{b}{}^{d} - u_{b}u^{d} \right)\tilde{\nabla}_{c}e_{d}=\\
    =& \frac{1}{2}N_{ab}\tilde{\phi} - u_{a}u_{b}\tilde{\mathcal{A}},
\end{split}
\end{equation}
and the brackets indicate the jump of a generic geometric quantity $W$ across the hypersurface,
\begin{equation}
    \left[ W \right] = \left. W^{+}\right|_{\mathcal{S}} - \left. W^{-}\right|_{\mathcal{S}}.
\end{equation}
Following  \cite{Vignolo2024oea}, we recognize, however, that the above conditions are insufficient to guarantee a non-singular geometry. The additional conditions that are required in the case of $f(\mathcal{Q})$ gravity are 
\begin{equation}\label{eq:junction_conditions_L}
    \left[ L_{cb}{}^{a} \right]n_{d} - \left[ L_{db}{}^{a} \right]n_{c} = 0,
\end{equation}
\begin{equation}\label{eq:junction_conditions_Q}
    \left. f'' \right|_{\mathcal{S}} = 0 \quad \cup \quad \left. P^{c}{}_{ab}n_{c} \right|_{\mathcal{S}} =0 \quad \cup \quad \left[ \mathcal{Q} \right] = 0,
\end{equation}
which are derived directly from the field equations.

Suppose the conditions on the extrinsic curvature, the disformation tensor, and the nonmetricity are not satisfied. In that case, the geometry can be regular only if we assume the presence of a thin shell separating the two regions $\mathcal{M}^{+}$ and $\mathcal{M}^{-}$. Such a thin shell can be proven to possess an energy-momentum tensor of the form\footnote{This is not the only possibility, though. Others are explored in \cite{Vignolo2024oea}.}
\begin{equation}\label{Shell_EMT}
\begin{split}
    \Psi^{\mathcal{S}}_{ab} =& \left(N_{ab} - u_{a}u_{b}\right)\left[\tilde{\mathcal{K}}\right] - \left[ \tilde{\mathcal{K}}_{ab} \right] +\\
    &- 2 \left.\frac{f''}{f'}P^{c}{}_{ab}n_{c}\right|_{\mathcal{S}}\left[\mathcal{Q}\right]=\\
    =&-\left[\tilde{\phi}\right]u_{a}u_{b} +\left(\frac{1}{2}\left[\tilde{\phi}\right] + \left[\tilde{\mathcal{A}} \right]\right)N_{ab} +\\ 
    &- 2 \left.\frac{f''}{f'}P^{c}{}_{ab}n_{c}\right|_{\mathcal{S}}\left[\mathcal{Q}\right],
\end{split}
\end{equation}
and the following curvature
\begin{equation}\label{Shell_Riemann}
\begin{split}
    \left({R^{\mathcal{S}}}\right)^{a}{}_{bcd} =& \left[ \Gamma_{cb}{}^{a} \right]n_{d} - \left[ \Gamma_{db}{}^{a} \right]n_{c}.
\end{split}
\end{equation}
Let us now apply these relations to our specific case. First, it follows from Eq. \eqref{eq:junction_conditions_Q} that a condition for ensuring a smooth junction is that the value of $\mathcal{Q}$ be the same on the boundary. Since we have seen that a Schwarzschild solution is characterized by the relation \eqref{eq:SdS_condition}, we can conclude that the junction can be satisfied for $\mathcal{Q}_{*}=0$. With this assumption, we have
\begin{equation}
    \Lambda_{*}=\frac{1}{2}\frac{f_0}{f'_0},
\end{equation}  
with $f_0=f(0)$ and $f'_0=f'(0)$. Therefore, the effective cosmological constant depends on the value of the function $f$ and its derivative in zero. At this point, it is necessary to observe that two functions $f(\mathcal{Q})$ and $g(\mathcal{Q})$, which differ by a constant term, can be considered the ``same'' theory of gravity. This is no different than considering GR as the same theory of gravity in the presence or absence of a cosmological constant. Consequently, it is meaningful to perform the junction between a solution of $f(\mathcal{Q})$ and $g(\mathcal{Q})$. Alternatively, one can write 
\begin{equation}
    f(\mathcal{Q})= 2\Lambda_{*}f'_0 + g(\mathcal{Q}),
\end{equation}
where $g(0)=0$, and therefore consider the constant term as a cosmological constant, albeit of non-metric origin, present in a theory with Lagrangian proportional to $g(\mathcal{Q})$.

With this idea in mind, we can apply the junction conditions. We assume that in the region $\mathcal{M}^{-}$ we have $\Lambda_{*}\neq0$  and that $\Lambda_{*}=0$ in $\mathcal{M}^{+}$. Therefore, we have the de Sitter solution in $\mathcal{M}^{-}$,
\begin{equation}
\begin{split}
    \left(\text{d}s^{2}\right)^{-} =& g_{ab} dX^a dX^b=- \left(1 - \frac{1}{3}\Lambda_{*} R^{2}\right)\text{d}T^{2} \\
    & + \left(1 - \frac{1}{3}\Lambda_{*} R^{2}\right)^{-1}\text{d}R^{2} \\
    &+ R^{2} \left( \text{d}\Theta^{2} + \sin^{2}{\Theta}\text{d}\varphi^{2}\right),   
\end{split}
\end{equation}
and the Schwarzschild one in $\mathcal{M}^{+}$,
\begin{equation}
\begin{split}
    \left(\text{d}s^{2}\right)^{+} =& g_{ab} dx^a dx^b=\\
    =& - \left(1 - \frac{r_{s}}{r}\right)\text{d}t^{2} + \left(1 - \frac{r_{s}}{r}\right)^{-1}\text{d}r^{2} \\
    &+ r^{2} \left( \text{d}\Theta^{2} + \sin^{2}{\Theta}\text{d}\varphi^{2}\right).
\end{split}
\end{equation}
Notice that in writing these metrics, we have already aligned the angular coordinates. This is possible because both metrics are spherically symmetric. Assuming $e_a$ as the normal vector $n_{a}$ to the hypersurface $\mathcal{S}$ in de Sitter spacetime, we have 
\begin{equation}\label{eq:normal_M-}
    n_{a}^{-} dX^a = e_{a}^{-} dX^a=\left(1 - \frac{1}{3}\Lambda_{*} R_{*}^{2}\right)^{-\frac{1}{2}} d R,
\end{equation}
in $\mathcal{M}^{-}$. With $R_{*}$ and $r_{*}$, we represent the position of $\mathcal{S}$ in the two different coordinate systems. To obtain the normal to $\mathcal{S}$ in $\mathcal{M}^{+}$, we perform the following coordinates transformation in Eq. \eqref{eq:normal_M-}: $t:T\longrightarrow t(T)$ and $r:R\longrightarrow r(R)$. Because of the normalization condition, which gives the relation
\begin{equation}\label{eq:norm_normal_M+}
    \left.\frac{\text{d} R}{\text{d} r} \right|_{\mathcal{S}} = \left(1 - \frac{1}{3}\Lambda_{*} R_{*}^{2}\right)^{\frac{1}{2}}\left(1 - \frac{r_{s}}{r_{*}}\right)^{-\frac{1}{2}},
\end{equation}
the result is that the normal in the exterior spacetime is equal to
\begin{equation}\label{eq:normal_M+}
    n_{a}^{+}dx^{a} = e_{a}^{+}dx^{a} = \left(1 - \frac{r_{s}}{r_{*}}\right)^{-\frac{1}{2}} d r.
\end{equation}
On the other hand, by using the orthogonality between $e^{a}$ and $u^{a}$, we obtain the following relations for the $4$-velocity:
\begin{equation}\label{eq:velocity_M-}
    u_{a}^{-} dX^{a}=\left(1 - \frac{1}{3}\Lambda_{*} R_{*}^{2}\right)^{\frac{1}{2}} d T,
\end{equation}
\begin{equation}\label{eq:velocity_M+}
    u_{a}^{+} dx^{a}=\left(1 - \frac{r_{s}}{r_{*}}\right)^{\frac{1}{2}} d t.
\end{equation}
Therefore, the induced metrics are equal to
\begin{equation}
\begin{split}
    q^{-}_{ab} dX^{a} dX^{b}=&(g^{-}_{ab}-e^{-}_{a} e^{-}_{b}) dX^{a} dX^{b}=\\
    =&- \left(1 - \frac{1}{3}\Lambda_{*} R_{*}^{2}\right)\text{d}T^{2} + \\
    &+ R_{*}^{2} \left( \text{d}\Theta^{2} + \sin^{2}{\Theta}\text{d}\varphi^{2}\right),   
\end{split}
\end{equation}
\begin{equation}
\begin{split}
    q^{+}_{ab} dx^{a} dx^{b}=&(g^{+}_{ab}-e^{+}_{a} e^{+}_{b}) dx^{a} dx^{b}\\
    =& - \left(1 - \frac{r_{s}}{r_{*}}\right)\text{d}t^{2} + \\
    &+ r_{*}^{2} \left( \text{d}\Theta^{2} + \sin^{2}{\Theta}\text{d}\varphi^{2}\right) =\\
    =& - \left(1 - \frac{1}{3}\Lambda_{*} R_{*}^{2}\right)\text{d}T^{2} + \\
     &+ r_{*}^{2} \left( \text{d}\Theta^{2} + \sin^{2}{\Theta}\text{d}\varphi^{2}\right).
\end{split}
\end{equation}
The junction condition \eqref{eq:junction_conditions_q} gives us that
\begin{equation}
    R_{*} = r_{*}.
\end{equation}

Using these results, we can verify that in $r_{*}$ Eq. \eqref{eq:junction_conditions_K} is not satisfied. We have 
\begin{equation}
\begin{split}
    \left[\tilde{\phi}\right]&=\frac{2 \sqrt{r_{*}-r_s}}{r_{*}^{3/2}}-\frac{2 \sqrt{1-\frac{1}{3}\Lambda _{*}
    r_{*}^{2}}}{r_{*}},\\
    \left[\tilde{\mathcal{A}} \right]&=\frac{r_s}{2 r_{*}^{3/2} \sqrt{r_{*}-r_s}}+\frac{\Lambda_{*} 
    r_{*}}{\sqrt{9-3\Lambda_{*}  r_{*}^{2}}},
\end{split}
\end{equation}
which cannot be put simultaneously to zero. Thus, a regular geometry can only be recovered by assuming the presence of a thin shell of matter separating the two regions. The stress-energy tensor of this form of matter can be deduced by the formula \eqref{Shell_EMT} and has
\begin{equation}\label{eq:mu_p_shell}
\begin{split}
     \rho^{\mathcal{S}}=&\Psi^{\mathcal{S}}_{ab}u^a u^b= \frac{2 \sqrt{1-\frac{1}{3}\Lambda_{*} r_{*}^{2}}}{r_{*}}-\frac{2 \sqrt{r_{*}-r_s}}{r_{*}^{3/2}},\\
     p^{\mathcal{S}}=&\frac{1}{2}\Psi^{\mathcal{S}}_{ab}N^{ab}=\frac{2 \Lambda_{*}  r_{*}^{2} - 3}{r_{*} \sqrt{9 - 3\Lambda_{*}  r_{*}^{2}}} + \frac{2r_{*} - r_{s}}{2 r_{*}^{3/2} \sqrt{r_{*}-r_s}}.
\end{split}
\end{equation}
The above quantities are positive or equal to zero for
\begin{equation}\label{eq:lambda_r_conditions}
 0 < \Lambda_{*}\leq 3\frac{r_{s}}{r_{*}^{3}}, \qquad  0 < r_{s} < r_{*}.
\end{equation}
As $r_{*}$ lies beyond the Schwarzschild radius, we can conclude that the above version of the Q-gravastar, despite its simplicity, is a consistent model for black hole alternatives in the context of $f(\mathcal{Q})$ gravity.

It is worth noting that for $\rho^{\mathcal{S}}=0$, making explicit the value of the Schwarzschild radius
\begin{equation}
    r_{s} = \frac{M}{4\pi},
\end{equation}
from the Eq. \eqref{eq:mu_p_shell} we have that the mass $M$ is equal to
\begin{equation}
    M = \frac{4}{3}\pi \Lambda_{*} r_{*}^{3},
\end{equation}
so the Schwarzschild metric is generated by a dark energy-filled sphere of radius $r_{*}$. On the other hand, for $\rho^{\mathcal{S}} \neq 0$, we obtain 
\begin{equation}
    M = \frac{4}{3}\pi  r_{*}^{3} \left[ \Lambda_{*} + \rho^{\mathcal{S}}\left(\frac{1}{r_{*}} \sqrt{9-3\Lambda_{*}  r_{*}^{2}} - \frac{3}{4} \rho^{\mathcal{S}} \right) \right],
\end{equation}
which is in agreement with the Eq. \eqref{eq:lambda_r_conditions} for
\begin{equation}
    0 < \rho^{\mathcal{S}} \leq 4 \sqrt{\frac{1}{r_{*}^{2}}-\frac{\Lambda_{*}}{3}}.
\end{equation}
Hence, in this case, the exterior spacetime is influenced by the energy density of the shell as well.


\section{Discussion and conclusions}\label{sec:conclusions}

We have developed the $1+1+2$ covariant formalism for LRS spacetimes of type II in the presence of nonmetricity. The resulting geometrical setting has been applied for studying the features of some homogeneous or static solutions arising in $f(\mathcal{Q})$ gravity, including the presence of Gravastar solutions.

We found that also in the case of $f(\mathcal{Q})$ gravity, the $1+1+2$ covariant formalism is very efficient in describing LRS spacetime and allows us to understand in a very detailed way the effect of nonmetricity, avoiding a specific choice of gauge. Special attention has been devoted to a complete decomposition of the Weyl tensor, which has revealed itself to be fundamental to properly analyzing spacetime dynamics.

Our analysis reveals that nonmetricity leads to substantial kinematic and dynamical differences compared to the purely metric case. For example, the timelike and spacelike congruences can be associated with two different types of acceleration, provided by Eqs. \eqref{eq:def_acceleration_u} and \eqref{eq:def_acceleration_e}, respectively. The difference between the accelerations is related to the geometric properties of the congruences themselves. One such property is autoparallelism.

As a first application of the general formalism, we have analyzed the case of homogeneous spacetimes, Sec. \ref{sec:hom_spacetime}. In particular, we specialized in studying the models whose spatial hypersurfaces are flat. Using the conditions derived from Eq. \eqref{eq:flatness_condition_on_3_Riemann}, we deduced several branches of solutions distinguished from the values assumed by the scalar components of the nonmetricity tensor. As an illustrative example, spatially flat FLRW spacetimes were used. Under appropriate assumptions on symmetry, we showed how to break covariance and find exact solutions of the metric and total connection.

A second application of the general formalism was the case of static and spherically symmetric spacetimes, Sec. \ref{sec:static_spacetime}. Using some simplifying assumptions on the nonmetricity tensor connected with the requirement of autoparallelism for the timelike congruence, we have deduced a self-consistent system of algebraic/differential equations for the investigation of vacuum solutions. These equations allow the identification of the conditions under which the theory produces the same Einstein-like field equations as GR. In addition, we have derived sufficient conditions that ensure the existence of a Schwarzschild-de Sitter solution in vacuum. Specifically, we found solutions characterized by the same metric but with different total connections.

In the context of spherically symmetric solutions, a further analysis concerns the presence of gravastars in the $f(\mathcal{Q})$ gravity. We have shown that one of the simplest possible gravastar solutions can be achieved by setting the nonmetricity scalar $\mathcal{Q}$ to zero. In this hypothesis, we have an effective cosmological constant, and the solution requires the presence of a shell that can be composed of standard matter. However, the equations indicate that this toy model is one possible realization of this scenario. Other realizations, for example, might lead to gravastars that do not require shells. In this respect, these solutions would be akin to Brill–Hartle Geons (see, e.g., \cite{brill1964method,anderson1997gravitational}).

The present work provides a new perspective on LRS solutions in $f(\mathcal{Q})$ gravity through the $1+1+2$ formalism. This perspective aims to lead to new insights in studying both cosmological and compact objects in non-metric theories of gravity.

\section*{Acknowledgments}
This work has been carried out in the framework of activities of the INFN Research Project QGSKY. 

\appendix

\section{Weyl tensor}\label{appendix_weyl}
In a metric theory, the Weyl tensor is defined by the Ricci decomposition as:
\begin{equation}
\begin{split}
    \tilde{C}_{abcd} =& \tilde{R}_{abcd} +\\
    &+\frac{1}{2}\left( g_{ad}\tilde{R}_{bc} - g_{ac}\tilde{R}_{bd} + g_{bc}\tilde{R}_{ad} - g_{bd}\tilde{R}_{ac} \right) +\\
    &+ \frac{1}{6}\tilde{R} \left( g_{ac}g_{bd}-g_{ad}g_{bc} \right).
\end{split}
\end{equation}
However, with nonmetricity, the Riemann tensor loses its antisymmetry in the first two indices, and therefore, we must also consider the homothetic curvature $\check{R}_{ab}$ \cite{McCrea:1992wa, JimenezCano:2021rlu}:
\begin{equation}
    R_{abcd} = W_{abcd} + Z_{abcd},
\end{equation}
with
\begin{equation}
    R_{[ab]cd} = W_{abcd} = \sum_{n=1}^{6} W^{(n)}_{abcd} ,
\end{equation}
\begin{equation}
    W^{(1)}_{abcd} = W_{abcd} - \sum_{n=2}^{6} W^{(n)}_{abcd} ,
\end{equation}
\begin{equation}
    W^{(2)}_{abcd} = \frac{1}{2}\left( W_{abcd} - W_{cdab} \right) - \left( R_{[mn]} - \bar{R}_{[mn]} \right) \delta^{m}_{\left[a\right.}g_{\left.b\right]\left[c\right.}\delta^{n}_{\left.d\right]},
\end{equation}
\begin{equation}
    W^{(3)}_{abcd} = W_{[abcd]} = R_{[abcd]},
\end{equation}
\begin{equation}
    W^{(4)}_{abcd} = - \left( R_{(mn)} - \bar{R}_{(mn)} \right) \delta^{m}_{\left[a\right.}g_{\left.b\right]\left[c\right.}\delta^{n}_{\left.d\right]} - \frac{1}{2}g_{c[a}g_{b]d}R,
\end{equation}
\begin{equation}
    W^{(5)}_{abcd} = \left( R_{[mn]} - \bar{R}_{[mn]} \right) \delta^{m}_{\left[a\right.}g_{\left.b\right]\left[c\right.}\delta^{n}_{\left.d\right]},
\end{equation}
\begin{equation}
    W^{(6)}_{abcd} = \frac{1}{6}g_{c[a}g_{b]d}R,
\end{equation}
\begin{equation}
    R_{(ab)cd} = Z_{abcd} = \sum_{n=1}^{5} Z^{(n)}_{abcd} ,
\end{equation}
\begin{equation}
    Z^{(1)}_{abcd} = Z_{abcd} - \sum_{n=2}^{5} Z^{(n)}_{abcd} ,
\end{equation}
\begin{equation}
\begin{split}
    Z^{(2)}_{abcd} =& \frac{1}{2}\left( Z_{abcd} - Z_{c(ab)d} + Z_{d(ab)c} \right) - \frac{1}{4}\left( R_{[mn]} +\right.\\ 
    &\left.+\bar{R}_{[mn]} - \check{R}_{mn} \right)\left( 2 \delta^{m}_{(a}g_{b)[c}\delta^{n}_{d]} - g_{ab}\delta^{m}_{[c}\delta^{n}_{d]} \right),
\end{split}
\end{equation}
\begin{equation}
\begin{split}
    Z^{(3)}_{abcd} =& \frac{1}{6}\left( R_{[mn]} + \bar{R}_{[mn]} +\right.\\ 
    &\left.- \frac{1}{2} \check{R}_{mn} \right)\left( 4 \delta^{m}_{(a}g_{b)[c}\delta^{n}_{d]} - g_{ab}\delta^{m}_{[c}\delta^{n}_{d]} \right),
\end{split}
\end{equation}
\begin{equation}
    Z^{(4)}_{abcd} = \frac{1}{2}\left( R_{(mn)} + \bar{R}_{(mn)} \right)\delta^{m}_{(a}g_{b)[c}\delta^{n}_{d]},
\end{equation}
\begin{equation}
    Z^{(5)}_{abcd} = \frac{1}{4}g_{ab}\check{R}_{cd}.
\end{equation}
We call ``general Weyl tensor'' the following expression,
\begin{equation}
    C_{abcd} = W^{(1)}_{abcd} + Z^{(1)}_{abcd}.
\end{equation}
It is straightforward to prove that $W^{(3)}_{abcd} = 0$ because of the first Bianchi identity \eqref{eq:first_bianchi_identity}.

\section{Tensor, vector and scalar elements of the magnetic and electric parts of the Weyl tensor}\label{appendix_magnetic_eletric_weyl}
We introduce the quantities involved in the Eqs. \eqref{eq:dec_magnetic_weyl_1}-\eqref{eq:dec_electric_weyl_3}:
\begin{equation}
    \mathcal{H}_{ab} = \left(N_{a}{}^{c}N_{b}{}^{d} - \frac{1}{2}N_{ab}N^{cd}\right)H_{cd},
\end{equation}
\begin{equation}
    \mathcal{H}_{a} = N_{a}{}^{c}e^{d}H_{cd}, \qquad \mathcal{H} = e^{a}e^{b} H_{ab} = - N^{ab}H_{ab}
\end{equation}
\begin{equation}
    \bar{\mathcal{H}}_{ab} = \left(N_{(a}{}^{c}N_{b)}{}^{d} - \frac{1}{2}N_{ab}N^{cd}\right)\bar{H}_{cd}, 
\end{equation}
\begin{equation}
    \bar{\mathcal{H}}_{a} = N_{a}{}^{c}e^{d}\bar{H}_{cd}, \qquad \mathbf{H}_{b} = e^{c}N_{b}{}^{d}\bar{H}_{cd}, 
\end{equation}
\indent
\begin{equation}
    \check{\bar{\mathcal{H}}}_{a} = N_{a}{}^{c}u^{d}\bar{H}_{cd}, \qquad
    \bar{\mathbb{H}} = \frac{1}{2}\epsilon^{ab}\bar{H}_{ab},
\end{equation}
\begin{equation}
    \check{\bar{\mathcal{H}}} = e^{a}u^{b}\bar{H}_{ab}, \qquad \bar{\mathcal{H}} = e^{a}e^{b} \bar{H}_{ab} = - N^{ab} \bar{H}_{ab},
\end{equation}
\begin{equation}
    \check{\mathcal{H}}_{ab} = \left(N_{(a}{}^{c}N_{b)}{}^{d} - \frac{1}{2}N_{ab}N^{cd}\right)\check{H}_{cd}, 
\end{equation}
\begin{equation}
    \check{\mathcal{H}}_{a} = N_{a}{}^{c}e^{d}\check{H}_{cd}, \qquad \check{\mathbf{H}}_{b} = e^{c}N_{b}{}^{d}\check{H}_{cd}, 
\end{equation}
\begin{equation}
    \mathfrak{H}_{a} = N_{a}{}^{c}u^{d}\check{H}_{cd} = \check{\bar{\mathcal{H}}}_{a}, \qquad
    \check{\mathbb{H}} = \frac{1}{2}\epsilon^{ab}\check{H}_{ab},
\end{equation}
\begin{equation}
    \mathfrak{H} = e^{a}u^{b}\check{H}_{ab} = \check{\bar{\mathcal{H}}}, \quad \check{\mathcal{H}} = e^{a}e^{b} \check{H}_{ab} = - N^{ab} \check{H}_{ab},
\end{equation}
\begin{equation}
    \mathcal{E}_{ab} = \left(N_{(a}{}^{c}N_{b)}{}^{d} - \frac{1}{2}N_{ab}N^{cd}\right)E_{cd}, 
\end{equation}
\begin{equation}
    \mathcal{E}_{a} = N_{a}{}^{c}e^{d}E_{cd}, \quad \mathbf{E}_{b} = e^{c}N_{b}{}^{d}E_{cd}, \quad \check{\mathcal{E}}_{b} = u^{c}N_{b}{}^{d}E_{cd},
\end{equation}
\begin{equation}
    \mathbb{E} = \frac{1}{2}\epsilon^{ab}E_{ab}, \qquad \check{\mathcal{E}} = u^{a}e^{b}E_{ab},
\end{equation}
\begin{equation}\label{eq:B10}
    \mathcal{E} = e^{a}e^{b} E_{ab} = - N^{ab} E_{ab},
\end{equation}
\begin{equation}
    \bar{\mathcal{E}}_{ab} = \left(N_{(a}{}^{c}N_{b)}{}^{d} - \frac{1}{2}N_{ab}N^{cd}\right)E_{cd}, 
\end{equation}
\begin{equation}
    \bar{\mathcal{E}}_{a} = N_{a}{}^{c}e^{d}E_{cd}, \quad
    \bar{\mathbf{E}}_{b} = e^{c}N_{b}{}^{d}E_{cd}, \quad \check{\bar{\mathcal{E}}}_{b} = u^{c}N_{b}{}^{d}E_{cd},
\end{equation}
\begin{equation}
    \bar{\mathbb{E}} = \frac{1}{2}\epsilon^{ab}\bar{E}_{ab}, \qquad \check{\bar{\mathcal{E}}} = u^{a}e^{b}\bar{E}_{ab} = \check{\mathcal{E}},
\end{equation}
\begin{equation}\label{eq:B14}
    \bar{\mathcal{E}} = e^{a}e^{b} \bar{E}_{ab} = - N^{ab} \bar{E}_{ab},
\end{equation}
\begin{equation}
    \mathfrak{E}_{ab} = N_{a}{}^{c}N_{b}{}^{d}\check{E}_{cd}, \qquad \mathfrak{E}_{a} = N_{a}{}^{c}e^{d}\check{E}_{cd},
\end{equation}
\begin{equation}
    \check{\mathfrak{E}}_{a} = N_{a}{}^{c}u^{d}\check{E}_{cd}, \qquad \mathfrak{E} = u^{c}e^{d}\check{E}_{cd} = \check{\mathcal{E}},
\end{equation}
\begin{equation}
    \check{\mathfrak{E}} = \frac{1}{2}\epsilon^{ab}\check{E}_{ab}.
\end{equation}

\section{Relations between torsion and vorticity}\label{appendix_torsion_vorticity}
Let us consider the intrinsic definition of the torsion tensor:
\begin{equation}\label{eq:torsion_def}
    T(x,y) = \nabla_{x}y - \nabla_{y}x + \left[ x , y \right],
\end{equation}
with $x$ and $y$ two generic vector fields. The same definition can be used to define the $3$-dimensional spatial torsion tensor, $\, ^{3}T_{ab}{}^{c}$, on the hypersurfaces we introduce by performing the $1+3$ splitting of spacetime:
\begin{equation}\label{eq:3d_torsion_def}
    \, ^{3}T(v,w) = D_{v}w - D_{w}v + \left[ v , w \right],
\end{equation}
where $v$ and $w$ are spatial vector fields,
\begin{equation}
    u_{a}v^{a} = u_{a}w^{a} = 0,
\end{equation}
and we have replaced the covariant derivative with the spatial one. 

In order to derive the relation between torsion and vorticity, we need to evaluate the following terms:\\
\begin{itemize}
    \item $\nabla_{v}w$,
    \begin{equation}\label{eq:torsion_dimon_1}
    \begin{split}
    &v^{a}\nabla_{a}w^{b} = v^{a} \delta_{a}{}^{c}\delta_{d}{}^{b}\nabla_{c}w^{d} =\\
    &= v^{a} \left( h_{a}{}^{c} - u_{a}u^{c} \right)\left( h_{d}{}^{b} - u_{d}u^{b} \right) \nabla_{c}w^{d} =\\
    &= v^{a}D_{a}w^{b} - v^{c}u_{d}u^{b}\nabla_{c}w^{d} = v^{a}D_{a}w^{b} + u^{a}w^{d}v^{c}\nabla_{c}u_{d} =\\
    &= v^{a}D_{a}w^{b} + u^{b}v^{c}w^{d}K_{cd};
    \end{split}
    \end{equation}
    \item $\nabla_{w}v$,
    \begin{equation}\label{eq:torsion_dimon_2}
    \begin{split}
    &w^{a}\nabla_{a}v^{b} = w^{a}D_{a}v^{b} + u^{b}v^{c}w^{d}K_{dc};
    \end{split}
    \end{equation}
\end{itemize}
If we substitute the Eqs. \eqref{eq:torsion_dimon_1} and \eqref{eq:torsion_dimon_2} in the Eq. \eqref{eq:3d_torsion_def}, and we consider that in our framework $T_{ab}{}^{c}$ is zero, we obtain
\begin{equation}\label{eq:3d_torsion_vorticity}
    \, ^{3}T_{ab}{}^{c} = - 2 \omega_{ab}u^{c},
\end{equation}
that is the  $3$-dimensional torsion tensor $\, ^{3}T_{ab}{}^{c}$  is proportional to the vorticity $\omega_{ab}$.

If we introduce the $2$-dimensional torsion tensor,
\begin{equation}
    \, ^{2}T(v,w) = \delta_{v}w - \delta_{w}v + \left[ v , w \right],
\end{equation}
a result similar to Eq. \eqref{eq:3d_torsion_vorticity} can be derived:
\begin{equation}
     \, ^{2}T_{ab}{}^{c} = - 2 \omega_{ab}u^{c} + 2 \epsilon_{ab} \xi e^{c},
\end{equation}
which also involves the vorticity $\xi$ that is related to the preferred spatial direction $e^{a}$.

\bibliographystyle{apsrev4-2}
\bibliography{LRS_spacetimes_of_type_II_in_f_Q_gravity.bib}

\end{document}